%% file: root.tex
\documentclass[twocolumn]{autart}    

\input{preamble.tex}

\begin{document}
\begin{frontmatter}

This paper has been published in Automatica (DOI: \href{https://doi.org/10.1016/j.automatica.2026.113031}{10.1016/j.automatica.2026.113031})

\title{Distributed consensus-based observer design \\for target state estimation with bearing measurements\thanksref{footnoteinfo}}

\thanks[footnoteinfo]{This paper was not presented at any conference. Corresponding author M.~Jacinto.}

\author[ISR]{Marcelo Jacinto}\ead{mjacinto@isr.tecnico.ulisboa.pt},
\author[ISR]{Pedro Trindade}\ead{pedro.trindade@isr.tecnico.ulisboa.pt},
\author[ISR,Copelabs]{Francisco Rego}\ead{frego@isr.tecnico.ulisboa.pt},
\author[ISR]{Rita Cunha}\ead{rita@isr.tecnico.ulisboa.pt},

\address[ISR]{Institute for Systems and Robotics, Instituto Superior T\'{e}cnico, Universidade de Lisboa, Portugal} 
\address[Copelabs]{Universidade Lus\'{o}fona, INESC INOV -- Lab, Lisboa, Portugal}

\input{Sections/Abstract}


\end{frontmatter}
\endNoHyper

\input{Sections/1_Introduction}

\input{Sections/2_Preliminaries}

\input{Sections/3_ProblemFormulation}

\input{Sections/4_DistributedObserverDesign}

\input{Sections/5_DistributedTargetTracking}

\input{Sections/6_NumericalResults}

\input{Sections/7_Conclusion}

\begin{ack}
The work of M. Jacinto and P. Trindade was supported by the PhD Grants (DOI: \href{https://doi.org/10.54499/2022.09587.BD}{10.54499/2022.09587.BD}) and (DOI: \href{https://doi.org/10.54499/2020.04749.BD}{10.54499/2020.04749.BD}) from Funda\c{c}{\~a}o para a Ci{\^e}ncia e a Tecnologia (FCT), Portugal. This work was also supported by FCT, Portugal through LARSyS (DOI: \href{https://doi.org/10.54499/LA/P/0083/2020}{10.54499/LA/P/0083/2020} and \href{https://doi.org/10.54499/UID/50009/2025}{10.54499/UID/50009/2025}). The work of Francisco Rego was funded through FCT, I.P project grants (DOI: \href{https://doi.org/10.54499/UID/06486/2025}{10.54499/UID/06486/2025}), (DOI: \href{https://doi.org/10.54499/UID/PRR/06486/2025}{10.54499/UID/PRR/06486/2025}), support grant (DOI: \href{https://doi.org/10.54499/UID/PRR2/06486/2025}{10.54499/UID/PRR2/06486/2025}), by COFAC/ILIND/COPELABS through the Seed Funding Program (7th edition) and project LoRaMAR (DOI: \href{https://doi.org/10.62658/COFAC/ILIND/COPELABS/1/2025}{10.62658/COFAC/ILIND/COPELABS/1/2025}). The authors gratefully acknowledge J. Pinto, G. Serrano, the reviewers and editors of the original manuscript for their suggestions that helped improve the quality of this work.
\end{ack}

\appendix
\input{Sections/Appendix/SchurLemma}
\input{Sections/Appendix/StabilityAnalysis}

\bibliographystyle{agsm}
\bibliography{bibliography}

\end{document}

%% file: preamble.tex
\usepackage{xcolor}
\usepackage{amssymb}
\usepackage{mathtools}
\usepackage{subcaption}
\usepackage{setspace}
\usepackage{siunitx}
\usepackage{natbib}
\usepackage{enumitem}
\usepackage{hyphenat}

\usepackage[bookmarks=false]{hyperref}
\hypersetup{hidelinks,
    linkcolor=green,  
    anchorcolor=black,
    citecolor=red,  
    filecolor=black,  
    menucolor=black,  
    urlcolor=blue,
    pdftitle={Distributed consensus-based observer design for target state estimation with bearing measurements},
    pdfauthor={Marcelo Jacinto, Pedro Trindade, Francisco Rego and Rita Cunha},
    pdfsubject={Automatica},	
    pdfkeywords={Multi-agent systems; Distributed localization; Bearing measurements; Consensus; Sensor networks.},
    pdfstartview=FitV,
    pdfdisplaydoctitle=true,
    breaklinks=true}
    
\newcommand{\vect}[1]{\mathbf{#1}}
\newcommand{\vectg}[1]{\boldsymbol{#1}}
\DeclareMathOperator{\diag}{diag}

\newtheorem{theorem}{Theorem}

\newtheorem{lemma}{Lemma}
\newtheorem{corollary}{Corollary}
\newtheorem{definition}{Definition}
\newtheorem{remark}{Remark}
\newtheorem{assumption}{Assumption}


%% file: Sections/Abstract.tex
\begin{abstract}
This paper introduces a novel distributed consensus-based observer design that enables a group of agents in an undirected communication network to solve the problem of target tracking, where the target is modelled as a chain of integrators of arbitrary order. Each agent is assumed to know its own position and simultaneously measure bearing vectors relative to the target. We start by introducing a general continuous time observer design tailored to systems whose state dynamics are modelled as chains of integrators and whose measurement model follows a particular nonlinear but observer-suited form. This design leverages a correction term that combines innovation and consensus components, allowing each agent to broadcast only a part of the state estimate to its neighbours, which effectively reduces the data flowing across the network. To provide uniform global exponential stability guarantees, a novel result for a class of nonlinear closed-loop systems in a generalized observer form is introduced and subsequently used as the main tool to derive stability conditions on the observer gains. Then, by exploring the properties of orthogonal projection matrices, the proposed design is used to solve the distributed target tracking problem and provide explicit stability conditions that depend on the target-agents geometric formation. Practical examples are derived for a target modelled as first-, second-, and third-order integrator dynamics, highlighting the design procedure and the stability conditions imposed. Finally, numerical results showcase the properties of the proposed algorithm.
\end{abstract}

%% file: Sections/1_Introduction.tex
\section{Introduction}
\label{section:Introduction}
Target localization and state estimation have long been an active area of research in the control community due to its wide range of practical applications in surveillance, aerial cinematography, and multi-vehicle formation tracking tasks (see for example \cite{Shao02012018,fei_gao,helical_guidance,ztang_observer_based_control_multi_vehicle_systems} and the references cited therein). In such application scenarios, it is commonplace to have agents equipped with onboard passive sensors, such as monocular cameras, that can provide relative position measurements to target features, modelled as unit bearing vectors (\cite{ducted_fan_le_bras}). However, when only a single bearing measurement to the target is available, the relative distance to the target cannot be directly recovered in general, and must be estimated instead. To address this estimation problem in single agent scenarios, several estimation frameworks have been proposed in the literature that assume a constant position (or velocity) motion model for the target, with a combination of  nonlinear or Linear Time-Varying (LTV) measurement models. In \cite{FARINA199961}, the bearing measurement is decomposed in a nonlinear function of elevation and azimuth angles and a Maximum Likelihood Estimator (MLE) framework is used to derive an observer for the target position. \cite{batista_2011_cdc} avoid having to explicitly linearize the measurement model. Instead, they rely on a Kalman Filter (KF) to solve the estimation problem by augmenting the system state with the relative range to the target, such that the nonlinear bearing measurement model can be expressed as an LTV quantity of the state. An alternative to augmenting the system state, is to introduce an orthogonal projection matrix that converts the nonlinear bearing measurement into a more tractable time-varying function of the target position (see \cite{5717795, BATISTA20131065}). For single agent target tracking, all the proposed approaches rely on Persistence of Excitation (PE) arguments, i.e., the relative motion between the agent and the target must have sufficient variation to ensure the observability of the system. In a multi-agent scenario, where multiple bearings are measured simultaneously by a network of agents, these PE conditions can, in general, be relaxed in favour of more practical geometric conditions on the agents' spatial distribution, i.e., a proper distribution of agents with respect to the target may be enough to ensure the observability of the system without requiring an explicit relative motion between the agents and the target (\cite{2024_ACC,ZHENG2025112216}). However, solving estimation problems in multi-agent systems also presents a set of key challenges, such as fusing the information gathered by all the agents without relying on a centralized filter or observer.

In the literature, there are many algorithms addressing the general problem of distributed estimation. In particular, \cite{REGO201936} provide a comprehensive survey on the topic of distributed estimation for discrete Linear Time-Invariant (LTI) systems, covering different Distributed Kalman Filter (DKF) methods, as well as consensus-based distributed Luenberger inspired methods, with many of the strategies highlighted in the survey also being generalizable to systems with time-varying measurements. One of the biggest challenges of DKF-based algorithms is the distributed computation of the filter covariance in an optimal and efficient manner. A common strategy to address this problem is to resort to covariance intersection methods, which provide a conservative approximation of the system's state covariance. \cite{battistelli_2015} summarize different variations of the DKF, centered around the covariance intersection method with consensus on information, and consensus on measurements. Although these methods tend to exhibit fast convergence rates, one of their main drawbacks is that they require a lot of information to flow across the network, as each agent is required to broadcast an information pair to its neighbours, composed of a state vector and an information- or covariance-related matrix for each consensus iteration. Moreover, they require multiple consensus iterations to be performed for every correction step executed by the filter. \cite{REGO2023111117} proposes an alternative design for LTV systems, where a processed version of the agents' measurements is shared across the network, instead of an information matrix, which for some system models requires the exchange of fewer data. An alternative to DKF methods are Distributed Recursive Least Squares (DRLS) algorithms, which are well suited to track slowly-varying processes (\cite{DRLS}). Another alternative to DKF methods is to consider distributed algorithms inspired by the Luenberger observer, which do not require the computation of covariance matrices and make use of fixed gains in the computation of the correction term. \cite{khan_et_al} propose a single-step consensus observer with fixed-gains for discrete time LTI systems, where the observer correction term combines consensus and innovation terms. In this method, both the estimated state and measurement vectors are shared among neighbouring agents, and stability guarantees are provided as a function of the observer gains, network connectivity, and the measurement model.

When considering the problem of target tracking in a distributed setting, practical constraints such as limited communication bandwidth can restrict the exchange of measurements or information matrices across the network of agents. This constraint motivates the development of Luenberger-inspired or adaptive distributed state observers, with fixed gains and embedded consensus terms, similar to those introduced by \cite{khan_et_al}. These observers typically minimize communication by broadcasting only the estimated state of the target to the network (see \cite{Shao02012018,DOU2020109022,Persistent_and_Intermittent_Bearing_Measurements}). Similar to their single agent counterparts, these distributed observers developed for target tracking with bearing measurements usually rely on orthogonal projection matrices and provide explicit stability guarantees by resorting to PE arguments. Alternatively, \cite{2024_ACC} and \cite{ZHENG2025112216}, make use of explicit spatial excitation arguments to obtain exponential stability guarantees that depend only on the geometric configuration of the agents relative to the target. In particular, \cite{ZHENG2025112216} propose a DRLS-based method to estimate the state of targets moving according to more complex motion models, but require each agent to broadcast to its neighbours the estimated state of the target as well as the corresponding bearing measurements and an observation matrix. \cite{2024_ACC} propose a continuous time nonlinear observer that only requires each agent to broadcast the estimated position of the target to the network, under the assumption that its position is stationary or slowly-drifting. This requirement highlights the limitations of existing distributed nonlinear observer frameworks -- they usually require a small amount of data to be exchanged across the network, but at the cost of assuming simple motion models for the target, which presents a significant challenge when tracking targets executing trajectories with complex high-order dynamics. There is also a lack of systematic design clues for extending these methodologies to handle scenarios where targets are modelled by more general models, such as higher-order integrators.

The main contribution of this work is a systematic distributed observer design to solve the target state estimation problem, where the target motion is modelled as a chain of integrators of arbitrary order. We consider a group of agents in an undirected communication network, where each agent is assumed to know its own position and simultaneously measure bearing vectors relative to the target. The proposed approach overcomes the limitations of existing methods, which are restricted to simpler target motion models. To that end, this work starts by introducing a general distributed observer framework with fixed-gains for chain-of-integrator systems in continuous time, with a particular class of nonlinear measurement models. Leveraging the integrator structure in the design of a correction term that combines innovation and consensus components, our approach reduces the amount of data exchanged between agents compared to other methods, proposed by \cite{battistelli_2015} and \cite{ZHENG2025112216}. Each agent broadcasts only the first element of the estimated state to neighbouring agents, and does not require the measurements to be shared across agents. To provide Uniform Global Exponential Stability (UGES) guarantees, a novel stability result is also introduced for closed-loop systems expressed in a generalized observer form with a nonlinear component. Subsequently, this result is used as the main tool to derive stability conditions for the proposed distributed observer. Finally, the proposed observer structure combined with the properties of orthogonal projection matrices are leveraged to solve the distributed target tracking problem, providing explicit spatial excitation conditions for the target-agents formation that ensure the observer convergence. Robustness of the proposed solution to measurements losses is also analysed, and practical examples are derived for targets modelled with constant position, velocity and acceleration, highlighting the systematic design procedure as the order of the target's motion model increases. Numerical simulation results are presented to illustrate the interplay between the observer gains and the geometric conditions imposed on the target-agents formation, as well as the cost of increasing the motion model order.

The remainder of this work is organized as follows. Section \ref{section:Preliminaries} introduces the notation, as well as the necessary mathematical and graph theoretical background, and Section \ref{section:DistributedTargetStateEstimation} describes the problem addressed in this paper. Section \ref{section:distributed_consensus_based_observer_design} presents a general observer design structure and the necessary tools for the stability analysis of the proposed framework. In Section \ref{section:distributed_target_tracking}, that observer is applied to the target state estimation problem (introduced in Section \ref{section:DistributedTargetStateEstimation}), and application examples are provided. Numerical results that illustrate the performance of the proposed algorithm are provided in Section \ref{section:SimulationResults}, and Section \ref{section:Conclusion} concludes the paper with final remarks.

%% file: Sections/2_Preliminaries.tex
\section{Notation and theoretical background}
\label{section:Preliminaries}
Throughout this work, vectors are lowercase bold while matrices are uppercase bold. The set of real numbers is denoted by $\mathbb{R}$, the subset of positive real numbers is denoted by $\mathbb{R}^{+}$, and the set of vectors in the 2-sphere is denoted by $\mathbb{S}^2$. The notation $\vect{S}^{N}$ is used to denote the set of symmetric $N \times N$ matrices, while $\vect{S}^{N}_{+}$ and $\vect{S}^{N}_{++}$ denote the subsets of positive semidefinite and positive definite matrices, respectively. The $N \times N$ identity matrix is denoted $\vect{I}_{N}$, while $\vect{0}$ and $\vect{1}$ denote vectors of zeros or ones with appropriate dimensions, respectively. The symbol $\otimes$ denotes the Kronecker product. The symbol $\|\cdot\|$ denotes the spectral norm when applied to matrices and $l_{2}$-norm when applied to vectors. The numeral superscript in parenthesis, i.e. $\vect{v}^{(m)}$, denotes the $m$-th component of a state vector $\vect{v}$, such that $\vect{v} = [\vect{v}^{(0)}, \vect{v}^{(1)}, \hdots, \vect{v}^{(M-1)}]$, and $\dot{\vect{v}}^{(m)}$ corresponds to the time-derivative of the $m$-th component. A block-diagonal matrix with $N$ matrices $\vect{A}_i \in \mathbb{R}^{D \times D}$ for $i \in \mathcal{N}\coloneqq\{1,\hdots,N\}$ is given by $\diag(\vect{A}_1,\hdots, \vect{A}_N) \in \mathbb{R}^{ND \times ND}$. Given a set of matrices $\vect{A}_{ij}$ of appropriate dimensions, we define the matrix $\vect{A}$ by its block entries, according to $\vect{A} = [\vect{A}_{ij}]$. Given a symmetric square matrix $\vectg{\Lambda}$, $\lambda_{\text{max}}(\vectg{\Lambda})$ denotes the maximum eigenvalue of $\vectg{\Lambda}$ and $\lambda_{\text{min}}(\vectg{\Lambda})$ the minimum eigenvalue. For any $N \times N$ symmetric matrices $\vect{C}$ and $\vect{D}$, the matrix inequalities $\vect{C} \succeq \vect{D}$ or $\vect{C} \succ \vect{D}$ are equivalent to stating that $\vect{C} - \vect{D} \in \vect{S}_{+}^{N}$ or $\vect{C} - \vect{D} \in \vect{S}_{++}^{N}$, respectively. The orthogonal projection matrix $\vectg{\Pi}_\vect{y} \in \vect{S}_{+}^{3}$ which projects an arbitrary vector $\vect{x} \in \mathbb{R}^{3}$ onto the subspace orthogonal to $\vect{y}\in \mathbb{S}^2$ is given by $\vectg{\Pi}_\vect{y}\coloneqq\vect{I}_{3} - \vect{y}\vect{y}^\top$.

\subsection{Schur complement}
\label{section:schur_complement}
Consider now the following results for a generic symmetric matrix $\vect{M}$ of the form
\begin{equation*}
	\vect{M} = \begin{bmatrix} \vect{A} & \vect{B} \\ \vect{B}^\top & \vect{C} \end{bmatrix}.
\end{equation*}

\begin{lemma}[\cite{boyd1994linear}]
\label{lemma:schur_complement}
Consider the matrix $\vect{M}$. The following results hold:
\begin{enumerate}[label=(\roman*)]
    \item If $\vect{C} \succ 0$, then $\vect{M} \succ 0 \iff \vect{A} - \vect{B} \vect{C}^{-1} \vect{B}^{\top} \succ 0$;
    \item If $\vect{A} \succ 0$, then $\vect{M} \succ 0 \iff \vect{C} - \vect{B}^{\top} \vect{A}^{-1} \vect{B} \succ 0$.
\end{enumerate}
\end{lemma}

By leveraging a known lower-bound on the smallest eigenvalue of $\vect{C}$ and the Schur complement, it is still possible to ensure that $\vect{M} \succ 0$ without explicitly computing $\vect{C}^{-1}$. Consider the result that follows.
\begin{lemma}
\label{lemma:schur_convergence_rate}
If there exists $\gamma > 0$ such that $\vect{C} \succ \gamma \vect{I}$, then $\vect{A}-\frac{1}{\gamma}\vect{B}\vect{B}^{\top} \succ 0 \implies \vect{A} - \vect{B}\vect{C}^{-1}\vect{B}^{\top} \succ 0$.
\label{lemma:schur_inverse_trick}
\end{lemma}
\textbf{PROOF.} The proof is presented in \autoref{appendix:proof_lemma_schur_inverse_trick}.

\subsection{Graph theory for undirected networks}
\label{section:graph_theory}
This section introduces the concept of an undirected communication network that will be used as a basis for the distributed observer design.

Consider a group of $N \geq 2$ agents, with a communication topology that can be modelled by a weighted undirected graph $\mathcal{G}(\mathcal{V}, \mathcal{E})$ consisting of a set of $N$ vertices $\mathcal{V} = \{1, ..., N\}$, a set of undirected edges $\mathcal{E} \subseteq \mathcal{V} \times \mathcal{V}$, and a weighted adjacency matrix $\vect{A} = [a_{ij}]\in \mathbb{R}^{N \times N}$ such that $a_{ij} > 0$ if the edge that connects vertex $i$ to $j$ belongs to the graph, i.e., $(i,j) \in \mathcal{E}$, and $a_{ij}=0$ otherwise. Self-edges $(i,i)$ are not allowed, meaning that $a_{ii} = 0$.

\begin{definition}
    For an undirected graph $\mathcal{G}$, if $(i,j) \in \mathcal{E}$ then $(j,i) \in \mathcal{E}$. As such, the set of neighbouring vertices of a vertex $i$ is given by $\mathcal{N}_{i} \coloneqq \{j \in \mathcal{V}: (i,j) \in \mathcal{E}\}$ and it follows directly that if $j \in \mathcal{N}_i$ then $i \in \mathcal{N}_j$.
\end{definition}

\begin{definition}
    A walk in a graph is an ordered sequence of vertices, such that a pair of consecutive vertices is an edge of the graph. The graph $\mathcal{G}$ is connected if there exists a walk between any two distinct vertices $i,j \in \mathcal{V}$.
\end{definition}

\begin{definition}
    The Laplacian matrix associated with the undirected graph $\mathcal{G}$ is given by $\vect{L} = [l_{ij}] \in \mathbb{R}^{N \times N}$ and is defined such that $l_{ij} = -a_{ij}$ if $i\neq j$ and $l_{ii} = \sum_{j \in \mathcal{N}_{i}} a_{ij}$. 
\end{definition}

The following result presents relevant properties of the Laplacian matrix that are leveraged throughout this work.

\begin{lemma}[\cite{FB-LNS}] 
\label{lemma:Laplacian_matrix}
The Laplacian matrix $\vect{L}$ associated with an undirected graph $\mathcal{G}$ is symmetric, positive semidefinite, and only has real eigenvalues, i.e., $\vect{L} \in \vect{S}^{N}_{+}$. If $\mathcal{G}$ is also connected, then $\vect{L}$ has only one null eigenvalue associated with eigenvector $\vect{1} \in \mathbb{R}^{N}$.
\end{lemma}

Note that the symmetric Laplacian $\vect{L}$ , associated with an undirected and connected graph $\mathcal{G}$, can be expressed via eigenvalue decomposition as $\vect{L} = \vect{V}\vect{J}\vect{V}^\top$. Furthermore, it follows from \autoref{lemma:Laplacian_matrix} that, without loss of generality, the matrices $\vect{J} \in \mathbb{R}^{N \times N}$ and $\vect{V} \in \mathbb{R}^{N \times N}$ can be expressed as
\begin{equation}
    \vect{J} = \begin{bmatrix}0 & \vect{0}^\top \\ \vect{0} & \vect{\Lambda}\end{bmatrix} \succeq 0
    \text{ and }
    \vect{V} = \begin{bmatrix}\vect{v}_1 & \hdots & \vect{v}_N \end{bmatrix},
    \label{eqn:eigen_vecto_decomposition_consensus}
\end{equation}
where $\vect{\Lambda} \in \vect{S}_{++}^{N-1}$ is a diagonal matrix with the ordered positive eigenvalues of $\vect{L}$, and $\vect{v}_i \in \mathbb{R}^N$, $i=1,\hdots,N$, are the corresponding eigenvectors of $\vect{L}$. In particular, $\vect{v}_1 = \frac{1}{\sqrt{N}} \vect{1}$ and $\vect{v}_i^\top\vect{v}_1=0$, $i=2,\hdots,N$. Furthermore, define the matrix $\vect{U} \coloneqq [\vect{v}_2, \hdots, \vect{v}_N] \in \mathbb{R}^{N \times N-1}$, from which the following equality holds
\begin{equation} 
\label{eqn:consensus_equality_property}
    \vect{I}_N = \frac{1}{N}\vect{1}\vect{1}^\top + \vect{U}\vect{U}^\top.
\end{equation}

%% file: Sections/3_ProblemFormulation.tex
\vspace{-0.6cm}
\section{System model and problem formulation}
\label{section:DistributedTargetStateEstimation}
Consider a point-mass target in 3-D space with state given by its position and the corresponding $M-1$ time-derivatives, according to
\begin{equation*}
    \vect{x}_{\mathrm{T}} \coloneqq [\vect{x}_{\mathrm{T}}^{(0)\top}, \vect{x}^{(1)\top}_{\mathrm{T}}, \hdots, \vect{x}^{(M-1)\top}_{\mathrm{T}}]^\top \in \mathbb{R}^{3M},
\end{equation*}
where $\vect{x}_{\mathrm{T}}^{(0)} = \vect{p}_{\mathrm{T}} \in \mathbb{R}^3$ denotes the position of the target with respect to an inertial frame $\{I\}$, and $M \geq 1$. In this context, the $m$-th state component also corresponds to the $m$-th time-derivative of the target position. It is assumed that the motion of the target can be modelled by a continuous autonomous LTI system following a chain of $M$-order integrators, according to 
\begin{equation}
	\dot{\vect{x}}_{\mathrm{T}}^{(m)} \coloneqq
	\begin{cases}
		\vect{x}_{\mathrm{T}}^{(m+1)} &\text{, if } m < M-1\\
		\vect{0} &\text{, if } m = M-1
	\end{cases},
    \label{eqn:target_dynamics}
\end{equation}
for $m=0,\hdots,M-1$. Consider also a group of $N$ autonomous agents communicating over an undirected network. Each agent $i=1,\hdots,N$ knows its own position $\vect{p}_{i} \in \mathbb{R}^3$ in the inertial frame $\{\mathcal{I}\}$, and is equipped with an onboard sensor, such that it can measure a unit bearing $\vect{b}_{i} \in \mathbb{S}^2$ from itself to the target, expressed in the inertial frame $\{\mathcal{I}\}$ according to
\begin{equation}
    \vect{b}_{i} \coloneqq \frac{\vect{p}_\mathrm{T} -  \vect{p}_i}{\|\vect{p}_\mathrm{T} -  \vect{p}_i\|} \in  \mathbb{S}^2.
    \label{eqn:bearing_measurement}
\end{equation}

\begin{figure}
    \centering
    \includegraphics[width=0.45\textwidth]{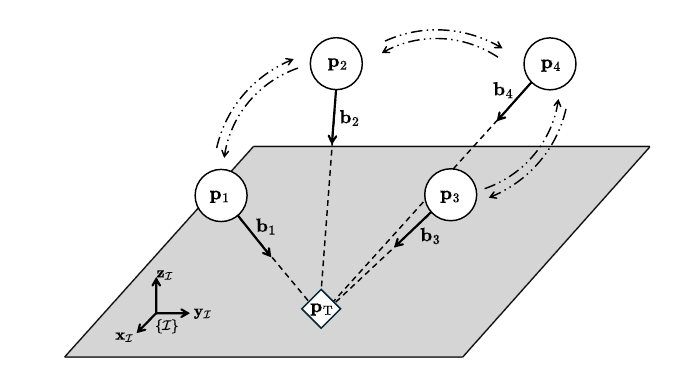}
    \caption{Distributed target state estimation example with four agents, where each agent measures a bearing vector, and exchanges the estimated target position with its neighbours.}
    \label{fig:problem_formulation_example}
\end{figure}

\vspace{-0.5cm}
The problem at hand consists in the development of a systematic design methodology for generating distributed observers with fixed-gains that enable the group of $N$ agents to estimate the full state $\vect{x}_{\mathrm{T}}$ of the single point-mass target, for any arbitrary $M \geq 1$, according to \autoref{fig:problem_formulation_example}.

This problem is addressed in two steps. First, a general distributed observer design methodology is proposed. Then, the nonlinear bearing measurement model is adapted so that the proposed design can be used to generate observers to estimate the target's state.

%% file: Sections/4_DistributedObserverDesign.tex
\vspace{-0.1cm}
\section{Distributed consensus-based observer design}
\label{section:distributed_consensus_based_observer_design}
This section introduces a systematic design methodology that can be used to generate continuous-time distributed observers for systems modelled as chains of an arbitrary number of integrators, with a particular class of nonlinear measurement models.

\vspace{0.8cm}
Consider a slightly more general autonomous system described by a chain of $M$-order integrators, such that
\begin{equation}
	\dot{\vect{x}}^{(m)}
	=
	\begin{cases}
		\vect{x}^{(m+1)} &\text{, if } m < M-1\\
		\vect{0} &\text{, if } m = M-1\\
	\end{cases},
	\label{eqn:general_system_dynamics_for_observer}
\end{equation}
for $m=0,\hdots,M-1$, where $\vect{x}^{(m)} \in \mathbb{R}^K$, and $\vect{x}\coloneqq[\vect{x}^{(0)\top}, \hdots, \vect{x}^{(M-1)\top}]^\top \in \mathbb{R}^{KM}$ with $K\geq 1$ and $M\geq 1$. Consider also a group of $N>1$ agents within a communication network, such that each agent acts both as a communication vertex and as a sensor capable of obtaining local measurements of the system, under the following assumptions.

\begin{assumption}
The communication topology of the group of $N$ agents is described by a fixed, undirected and connected graph $\mathcal{G}$, with an associated weighted adjacency matrix $\vect{A}$ and corresponding Laplacian matrix $\vect{L}$.
\label{assumption:undirected_communication_network}
\end{assumption}

\begin{assumption}
Each agent $i \in \mathcal{V}=\{1,\dots,N\}$ is equipped with a sensor that can be described by a nonlinear measurement model, according to
\begin{equation}
    \vect{y}_i \coloneqq \vectg{\Psi}_i(t, \vect{x}) \vect{x}^{(0)} \in \mathbb{R}^{K},
    \label{eqn:general_time_varying_measurement_model}
\end{equation}
where $\vectg{\Psi}_i(t, \vect{x}) \in \vect{S}^{K}$ is a continuous and bounded function of $t$ and $\vect{x}$.
\label{assumption:agent_measurement_model}
\end{assumption}

Notice that each agent is assumed to have a measurement model that depends explicitly on $\vect{x}^{(0)} \in \mathbb{R}^{K}$, but the observation matrix can be different for each individual agent. This allows us to recover the scenario presented by \cite{battistelli_2015}, in which some agents act as sensor vertices, and others only act as communication vertices by considering the pair $(\vectg{\Psi}_i, \vect{y}_i)=(\vect{0},\vect{0})$ if no measurements are available for some agent $i$. Further details on the minimum number of agents containing a measurement model is detailed in the sections that follow. Analogously, $\vectg{\Psi}_i=\vect{I}_K$ can be considered if the first component of the system state can be fully measured by agent $i$, although this last case is not the focus of this work.

\subsection{Distributed observer design}
Consider the system described by \eqref{eqn:general_system_dynamics_for_observer} with measurement model given by \eqref{eqn:general_time_varying_measurement_model}. Define $\hat{\vect{x}}_i^{(m)} \in \mathbb{R}^{K}$ as the estimate of the $m$-th component of the system state computed by agent $i$. The proposed  estimation algorithm for each agent $i$ is given by
\begin{equation}
	\dot{\hat{\vect{x}}}_i^{(m)} 
    \coloneqq
    \begin{cases}
    	\hat{\vect{x}}_i^{(m+1)} + k_{m+1} \vectg{\delta}_i &\text{, if } m < M-1\\
    	k_{M} \vectg{\delta}_i &\text{, if } m=M-1
    \end{cases}
    \label{eqn:distributed_observer}
\end{equation}
where $\vect{k} = [k_1, \hdots, k_M] \in \mathbb{R}^{M}$ is a constant gains vector. This model borrows inspiration from a Luenberger observer, as it follows a replica of the system dynamics with a correction term $\vectg{\delta}_i \in \mathbb{R}^{K}$. This correction term combines an innovation term and a first-order consensus term, and is defined according to
\begin{equation}
    \vectg{\delta}_i \coloneqq \vect{y}_i - \vectg{\Psi}_{i}(t, \vect{x})\hat{\vect{x}}_i^{(0)} - \alpha \sum_{j \in \mathcal{N}_{i}} a_{ij}(\hat{\vect{x}}_i^{(0)} - \hat{\vect{x}}_j^{(0)}),
    \label{eqn:distributed_observer_correction_term}
\end{equation}
with $\alpha \in \mathbb{R}^{+}$ a coupling gain, and $a_{ij} \in \mathbb{R}^{+}$ the corresponding term from the weighted adjacency matrix associated with network graph $\mathcal{G}$. The innovation term quantifies the mismatch between the expected and the actual observation of agent $i$. The consensus term weights the mismatch between the estimates of the first element of the state between neighbouring agents. 

The proposed correction term shares some similarity with the method presented by \cite{khan_et_al} for LTI systems, in the sense that it combines both an innovation and a consensus component in a single correction term. However, by leveraging the integrator structure in the system model, our proposal only requires the first element of the system state $\hat{\vect{x}}_{i}^{(0)} \in \mathbb{R}^{K}$ to be broadcast to neighbouring agents, and does not require the measurements to be shared across agents.

\subsection{Stability criteria for observer design}
Before proceeding with a formal stability analysis of the proposed observer design, we first introduce intermediate stability results for closed-loop systems in a generalized observer form. These results will then serve as auxiliary tools for the stability analysis of the proposed framework.

\begin{theorem}
\label{theorem:stability_of_nonlinear_system}
Consider a closed-loop system described by the state vector $\vectg{\xi} \coloneqq [\vectg{\xi}^{(0)}, \hdots, \vectg{\xi}^{(M-1)}]^{\top} \in \mathbb{R}^{WM}$, where $\vectg{\xi}^{(m)} \in \mathbb{R}^{W}$ is the $m$-th order system state, with $M\geq 1$, and system dynamics given by
\begin{equation}
	\dot{\vectg{\xi}} = \vectg{\Xi}(t,\vectg{\xi}) \vectg{\xi},
	\label{eqn:closed_loop_nonlinear_observer_canonical_form}
\end{equation}
with
\begin{equation}
	\vectg{\Xi}(t,\vectg{\xi})
	\coloneqq
	\begin{bmatrix}
        -k_1\vectg{\Phi}(t,\vectg{\xi}) & \vect{I}_W & \vect{0}  & \hdots & \vect{0} \\
        -k_2\vectg{\Phi}(t,\vectg{\xi}) & \vect{0}  & \vect{I}_W & \hdots & \vect{0} \\
        \vdots              & \vdots    & \vdots     & \ddots & \vdots   \\
        -k_{M-1}\vectg{\Phi}(t,\vectg{\xi}) & \vect{0}  & \vect{0} & \hdots & \vect{I}_W \\
        -k_M\vectg{\Phi}(t,\vectg{\xi}) & \vect{0}  & \vect{0}   & \hdots & \vect{0} \\
    \end{bmatrix},
    \label{eqn:time_variant_matrix_error_structure}
\end{equation}
where $\vectg{\Phi}(t,\vectg{\xi}) \in \vect{S}^{W}_{++}$ is continuous and bounded function of $t$ and $\vectg{\xi}$, and $k_m \in \mathbb{R}$, $m={1, \hdots, M}$, are positive system gains. The origin of the closed loop system \eqref{eqn:closed_loop_nonlinear_observer_canonical_form} is UGES if there exists $\delta > 0$, such that, for all $t \geq t_0$:
\begin{enumerate}[label=(\roman*)]
    \item the matrix $\vectg{\Phi}(t,\vectg{\xi})$ satisfies the inequality
    \begin{equation}
        \vectg{\Phi}(t,\vectg{\xi}) \succ \mu \vect{I}_W,
        \label{eqn:condition_on_the_phi_matrix}
    \end{equation}
    with $\mu = \delta$ for $M=1$ and $\mu = (\delta k_1 + k_2)/k_1^2$ for $M \geq 2$;
    \label{theorem:closed_loop:enumerate:item:first}
    \item for $M \geq 2$ the Linear Matrix Inequality (LMI) given by
    \begin{equation}
        \Bar{\vect{Q}} \coloneqq \bar{\vectg{\Sigma}} + \bar{\vectg{\Sigma}}^{\top} \succ 0
        \label{eqn:LMI_for_theorem_1}
    \end{equation}
    holds, where each entry of the matrix $\bar{\vectg{\Sigma}}$ is defined by ratios of the system gains $c_l \coloneqq k_{l+1} / k_{l}$ and by $\delta$, according to
    \begin{equation}
    	[\bar{\vectg{\Sigma}}_{ij}]
    	\! \coloneqq \!
    	\begin{cases}
        	c_{M-1} & \mathrm{, if } \: i=1\\
            c_{M-i}-c_{M-i+1} & \mathrm{, if } \: 2 \leq i \leq j < M\\
        	-c_{M-j} & \mathrm{, if } \: i=j+1\\
            \delta & \mathrm{, if } \: j=i=M \\
        	0 & \mathrm{, if } \: i>j+1 
    	\end{cases}.
    	\label{eqn:gains_ratio_definition_theorem}
    \end{equation}
    \end{enumerate}
    \label{theorem:closed_loop:enumerate:item:second}
\end{theorem}

\textbf{PROOF.} The proof is presented in \autoref{appendix:proof_theorem_stability_of_nonlinear_system}.

\begin{remark}
	Notice that for $M=2$, the LMI imposed by \eqref{eqn:LMI_for_theorem_1} is automatically satisfied for any choice of gains $k_1, k_2 > 0$ and $\delta > 0$, since $\bar{\vect{Q}} = 2\diag(k_{2} / k_{1}, \delta)$.
    \label{remark:M2_LMI_is_satistisfied}
\end{remark}

Building on this result, we extend the stability analysis to consider system \eqref{eqn:closed_loop_nonlinear_observer_canonical_form} subject to a bounded input. The following Lemma will be later used as an auxiliary tool to analyse the performance of the observer when there is a mismatch between the real target motion and the adopted motion model.
\vspace{-0.2cm}
\begin{lemma}
\label{lemma:ISS_stability_of_nonlinear_system}
Consider the system \eqref{eqn:closed_loop_nonlinear_observer_canonical_form} with a bounded and piecewise continuous input $\vect{u} \in \mathbb{R}^{W}$, defined according to
\begin{equation}
    \dot{\vectg{\xi}} = \vectg{\Xi}(t,\vectg{\xi}) \vectg{\xi} + \vect{B}\otimes \vect{u},
    \label{eqn:closed_loop_nonlinear_observer_canonical_form_with_input}
\end{equation}
with $\vect{B} = [\vect{0}^{\top} \, 1]^{\top} \in \mathbb{R}^{M}$. The system is input-to-state stable (ISS) with respect to the input $\vect{u}$, provided that assertions (i) and (ii) from \autoref{theorem:stability_of_nonlinear_system} are satisfied.
\end{lemma}
\textbf{PROOF.} The proof is presented in \autoref{appendix:proof_lemma_ISS_stability_of_nonlinear_system}.

Now that all the necessary ingredients for analysing the stability of the distributed observer have been introduced, we can proceed with the main result of this work.
\begin{theorem}
\label{theorem:theorem_distributed_consensus_observer}
Consider the autonomous system given by \eqref{eqn:general_system_dynamics_for_observer} and the distributed observer described by \eqref{eqn:distributed_observer} and \eqref{eqn:distributed_observer_correction_term}. Under Assumptions \ref{assumption:undirected_communication_network} and \ref{assumption:agent_measurement_model}, the error system associated with the distributed observer has a UGES equilibrium point at the origin if there exist $\delta, \gamma > 0$, such that for all $t > t_0$:
\begin{enumerate}[label=(\roman*)]
    \item the inequality given by
    \begin{equation}
            \frac{1}{N} \sum_{i=1}^{N} \vectg{\Psi}_i(t, \vect{x}) \succ (\mu + \gamma) \vect{I}_K,
        \label{eqn:generic_geometric_condition}
    \end{equation}
    holds with $\mu = \delta$ for $M=1$ or $\mu = (\delta k_1 + k_2)/k_1^2$ for $M \geq 2$;
    \item the consensus gain $\alpha$ satisfies
        \begin{equation}
            \alpha > \frac{\mu + \left \|\frac{1}{\gamma}\vectg{\Psi}^2(t,\vect{x}) - \vectg{\Psi}(t, \vect{x}) \right \|}{\lambda_{\min}(\vectg{\Lambda})},
        \label{eqn:general_consensus_gain_condition}
        \end{equation}
    where $\lambda_{\min}(\vectg{\Lambda})$ corresponds to the smallest positive eigenvalue of the Laplacian matrix $\vect{L}$, and 
    \begin{equation}
    	\vect{\Psi}(t, \vect{x}) \coloneqq \diag(\vectg{\Psi}_{1}, \hdots, \vectg{\Psi}_{N}) \in \vect{S}_{+}^{KN};
        \label{eqn:diagonal_psi}
    \end{equation}
    \item for $M \geq 2$, the observer gains $\vect{k}$ is chosen such that $\bar{\vect{Q}} \succ 0$, defined according to \eqref{eqn:LMI_for_theorem_1} and \eqref{eqn:gains_ratio_definition_theorem}.
\end{enumerate}
\end{theorem}
\textbf{PROOF.} Let the estimation error $\tilde{\vect{x}}_i^{(m)} \in \mathbb{R}^{K}$ for the $m$-order component of the state at agent $i$ be given by
\begin{equation}
    \tilde{\vect{x}}_i^{(m)} \coloneqq \vect{x}^{(m)} - \hat{\vect{x}}_i^{(m)},
    \label{eqn:general_error_by_derivative_order}
\end{equation}
with $m=0,\hdots, M-1$. Replacing \eqref{eqn:general_system_dynamics_for_observer}, \eqref{eqn:general_time_varying_measurement_model}, \eqref{eqn:distributed_observer} and \eqref{eqn:distributed_observer_correction_term} in the time-derivative of \eqref{eqn:general_error_by_derivative_order} yields the error dynamics
\begin{equation*}
\dot{\tilde{\vect{x}}}_i^{(m)} =
\begin{cases}
    \tilde{\vect{x}}_i^{(m+1)} - k_{m+1} \tilde{\vectg{\delta}}_i &\text{, if } m < M-1\\
    - k_{M} \tilde{\vectg{\delta}}_i &\text{, if } m=M-1
\end{cases},
\end{equation*}
where $\tilde{\vectg{\delta}}_i = \vectg{\Psi}_{i}(t, \vect{x})\tilde{\vect{x}}^{(0)}_i + \alpha \sum_{j \in \mathcal{N}_{i}} a_{ij}(\tilde{\vect{x}}_i^{(0)} - \tilde{\vect{x}}_j^{(0)})$.

We define the $m$-th order system error according to $\tilde{\vect{x}}^{(m)} \coloneqq [\tilde{\vect{x}}^{(m)\top}_1, \hdots, \tilde{\vect{x}}^{(m)\top}_N]^\top \in \mathbb{R}^{KN}$, and $\tilde{\vect{x}} \coloneqq [\tilde{\vect{x}}^{(0)\top}, \hdots, \tilde{\vect{x}}^{(M-1)\top}]^{\top} \in \mathbb{R}^{KNM}$. Then, the total system error dynamics can be expressed according to
\begin{equation}
    \begin{bmatrix}
        \dot{\tilde{\vect{x}}}^{(0)} \\
        \dot{\tilde{\vect{x}}}^{(1)} \\
        \vdots \\
        \dot{\tilde{\vect{x}}}^{(M-1)}
    \end{bmatrix}
    \!\! = \!\!
    \begin{bmatrix}
        -k_1\vectg{\Phi}(t,\tilde{\vect{x}}) & \vect{I}_{KN} & \vect{0}  & \hdots & \vect{0} \\
        -k_2\vectg{\Phi}(t,\tilde{\vect{x}}) & \vect{0}  & \vect{I}_{KN} & \hdots & \vect{0} \\
        \vdots              & \vdots    & \vdots     & \ddots & \vdots   \\
        -k_M\vectg{\Phi}(t,\tilde{\vect{x}}) & \vect{0}  & \vect{0}   & \hdots & \vect{0} \\
    \end{bmatrix}
    \!\!
    \begin{bmatrix}
        \tilde{\vect{x}}^{(0)} \\
        \tilde{\vect{x}}^{(1)} \\
        \vdots \\
        \tilde{\vect{x}}^{(M-1)}
    \end{bmatrix},
    \label{eqn:total_closed_loop_observer_error_dynamics}
\end{equation}
with
\vspace{-0.2cm}
\begin{equation}
    \vectg{\Phi}(t,\tilde{\vect{x}}) \coloneqq \vectg{\Psi}(t,\vect{x}) + \alpha (\vect{L} \otimes \vect{I}_K) \in \vect{S}_{+}^{KN},
    \label{eqn:theorem:definition_of_phi_observer}
\end{equation}
where $\vectg{\Psi}(t,\vect{x})$ is the block-diagonal matrix defined according to \eqref{eqn:diagonal_psi}, and the relationship between $\vect{x}$ and $\tilde{\vect{x}}$ is given by \eqref{eqn:general_error_by_derivative_order}. Hereafter, when clear from context, the explicit dependence on time and the state vector is dropped.

\vspace{0.5cm}
Consider that each element $\tilde{\vect{x}}^{(m)} \in \mathbb{R}^{KN}$ corresponds to $\vectg{\xi}^{(m)} \in \mathbb{R}^{W}$ in \eqref{eqn:closed_loop_nonlinear_observer_canonical_form}, with $W=KN$. By direct application of \autoref{theorem:stability_of_nonlinear_system} and \autoref{remark:M2_LMI_is_satistisfied}, it follows that the origin of the estimation error vector is UGES if \textit{i)}~there exists $\delta > 0$ such that for all $t \geq t_0$
\begin{equation}
    \vectg{\Phi} \succ \mu \vect{I}_{KN},
    \label{eqn:proof_condition_of_time_varying_matrix}
\end{equation}
with $\mu=\delta$ for $M=1$ or $\mu = (\delta k_1 + k_2)/k_1^2$ for $M \geq 2$, and \textit{ii)}~the LMI $\Bar{\vect{Q}} \succ 0$ is satisfied for $M \geq 2$, with matrix $\bar{\vect{Q}}$ defined according to \eqref{eqn:LMI_for_theorem_1} and \eqref{eqn:gains_ratio_definition_theorem}. These conditions are sufficient to guarantee the stability of the distributed observer. 

\vspace{-0.3cm}
To explore \eqref{eqn:proof_condition_of_time_varying_matrix} for use in the design of stabilizing observer gains, consider a similarity transformation that preserves the definiteness of $\vectg{\Phi}-\mu\vect{I}_{KN} \succ 0$, according to 
\begin{equation*}
    \bar{\vectg{\Phi}} \coloneqq (\vect{V}^{\top} \otimes \vect{I}_K)[\vectg{\Phi}-\mu \vect{I}_{KN}](\vect{V} \otimes \vect{I}_K),
\end{equation*}
where $\vect{V} \in \mathbb{R}^{N\times N}$, previously introduced in \eqref{eqn:eigen_vecto_decomposition_consensus}, denotes the matrix of eigenvectors associated with the graph Laplacian $\vect{L}$. The LMI $\bar{\vectg{\Phi}}(t, \vect{x}) \succ 0$ can be further decomposed according to
\begin{equation*}
    \vectg{\bar{\Phi}} = 
    \begin{bmatrix}
        \vectg{\bar{\Phi}}_{11} & \vectg{\bar{\Phi}}_{12} \\
        \vectg{\bar{\Phi}}^{\top}_{12} & \vectg{\bar{\Phi}}_{22}
    \end{bmatrix},
\end{equation*}
with
\begin{equation*}
\begin{split}
    \vectg{\bar{\Phi}}_{11} =& \frac{1}{N} (\vect{1}^{\top} \otimes \vect{I}_K) \vectg{\Psi} (\vect{1} \otimes \vect{I}_K) - \mu \vect{I}_K\\
    =& \frac{1}{N}\sum_{i=1}^{N}\vectg{\Psi}_i - \mu \vect{I}_K \in \vect{S}^{K},\\[8pt]
    \vectg{\bar{\Phi}}_{12} =& \frac{1}{\sqrt{N}} (\vect{1}^{\top} \otimes \vect{I}_K) \vectg{\Psi} (\vect{U} \otimes \vect{I}_K) \in \mathbb{R}^{K \times K(N-1)}, \\[8pt]
    \vectg{\bar{\Phi}}_{22} =& \alpha (\vectg{\Lambda} \otimes \vect{I}_K) - \mu \vect{I}_{K(N-1)} \\
    &+ (\vect{U}^{\top} \otimes \vect{I}_K) \vectg{\Psi} (\vect{U} \otimes \vect{I}_K) \in \vect{S}^{K(N-1)},
\end{split}
\end{equation*}
where $\vect{U} \in \mathbb{R}^{N \times N-1}$ denotes the matrix of eigenvectors associated with $\vect{\Lambda} \in \vect{S}_{++}^{N-1}$, the diagonal matrix with the positive eigenvalues of $\vect{L}$. From \autoref{lemma:schur_complement}, the matrix $\bar{\vectg{\Phi}}(t,\vect{x})$ is positive definite if and only if $\vectg{\bar{\Phi}}_{11}(t,\vect{x}) \succ 0$ and
\begin{equation}
    \vectg{\bar{\Phi}}_{22} - \vectg{\bar{\Phi}}_{12}^{\top}\vectg{\bar{\Phi}}_{11}^{-1}\vectg{\bar{\Phi}}_{12} \succ 0.
    \label{eqn:time_varying_component_LMI}
\end{equation}
Since it is not possible to pre-compute $\vectg{\bar{\Phi}}_{11}(t,\vect{x})$ explicitly for all $t \geq t_0$, consider the conservative assumption that $\vectg{\bar{\Phi}}_{11}(t,\vect{x})$ is lower-bounded such that $\vectg{\bar{\Phi}}_{11}(t,\vect{x}) \succ \gamma \vect{I}_K$, with $\gamma > 0$ for all $t \geq t_0$, i.e., the inequality \eqref{eqn:generic_geometric_condition} is satisfied. From direct application of \autoref{lemma:schur_inverse_trick}, it follows that inequality \eqref{eqn:time_varying_component_LMI} holds if 
\begin{equation*}
    \vectg{\bar{\Phi}}_{22} - \frac{1}{\gamma}\vectg{\bar{\Phi}}_{12}^{\top}\vectg{\bar{\Phi}}_{12} \succ 0.
\end{equation*}
Expanding the inequality yields
\begin{equation*}
\begin{split}
    &\alpha (\vectg{\Lambda} \otimes \vect{I}_K) \succ  \mu \vect{I}_{K} - (\vect{U}^{\top} \otimes \vect{I}_K) \vectg{\Psi} (\vect{U} \otimes \vect{I}_K)\\
    &+ \frac{1}{\gamma N} (\vect{U}^{\top} \otimes \vect{I}_K)\vectg{\Psi}(\vect{1}\vect{1}^{\top} \otimes \vect{I}_K)\vectg{\Psi}(\vect{U} \otimes \vect{I}_K),
\end{split}
\label{eqn:intermediate_lmi_general_proof}
\end{equation*}
and exploiting the property introduced in \eqref{eqn:consensus_equality_property}, it allows us to rewrite it as
\begin{equation}
\begin{split}
    \alpha &(\vectg{\Lambda} \otimes \vect{I}_K) \succ \mu \vect{I}_{K} \\
    &+(\vect{U}^{\top} \otimes \vect{I}_K)\left(\frac{1}{\gamma}\vectg{\Psi}^2 - \vectg{\Psi} \right)(\vect{U} \otimes \vect{I}_K)\\
    &-\frac{1}{\gamma} (\vect{U}^{\top} \otimes \vect{I}_K)\vectg{\Psi}(\vect{U}\vect{U}^{\top} \otimes \vect{I}_K)\vectg{\Psi}(\vect{U} \otimes \vect{I}_K).
\end{split}
\label{eqn:inequality_for_the_consensus_gain_proof}
\end{equation}
Notice that the last term of the inequality satisfies 
\begin{equation*}
    \frac{1}{\gamma} (\vect{U}^{\top} \otimes \vect{I}_K)\vectg{\Psi}(\vect{U}\vect{U}^{\top} \otimes \vect{I}_K)\vectg{\Psi}(\vect{U} \otimes \vect{I}_K) \succeq 0,
\end{equation*}
regardless of the definiteness of $\vectg{\Psi}$. Therefore, it follows directly that \eqref{eqn:inequality_for_the_consensus_gain_proof} is satisfied if the more conservative bound
\begin{equation*}
    \alpha (\vectg{\Lambda} \otimes \vect{I}_K) \succ \mu \vect{I}_{K} + (\vect{U}^{\top} \otimes \vect{I}_K)\left(\frac{1}{\gamma}\vectg{\Psi}^2 - \vectg{\Psi} \right)(\vect{U} \otimes \vect{I}_K)
\end{equation*}
also holds. Moreover, this bound is satisfied if the consensus gain $\alpha$ satisfies 
\begin{equation*}
    \alpha > \frac{\mu + \left\| (\vect{U}^{\top} \otimes \vect{I}_K)\left(\frac{1}{\gamma}\vectg{\Psi}^2 - \vectg{\Psi} \right)(\vect{U} \otimes \vect{I}_K) \right\|}{\lambda_{\min}(\vectg{\Lambda})}.
\end{equation*}
Using the fact that the columns of $\vect{U}$ are orthonormal and $\|(\vect{U} \otimes \vect{I}_{K})\| = 1$, we can finally conclude that the LMI given by \eqref{eqn:time_varying_component_LMI} is satisfied if the conservative bound \eqref{eqn:general_consensus_gain_condition} imposed on the consensus gain $\alpha$ holds. As such, \eqref{eqn:proof_condition_of_time_varying_matrix} is satisfied if both \eqref{eqn:generic_geometric_condition} and \eqref{eqn:general_consensus_gain_condition} hold. \qed

\begin{remark}
    In a scenario where some agents do not have measurements available, i.e., $(\vectg{\Psi}_i, \vect{y}_i)=(\vect{0},\vect{0})$, condition (i) will dictate how many agents in the network can act only as communication vertices while still ensuring UGES.
    \label{remark:no_measurements_available}
\end{remark}

The bounds $\mu$ and $\gamma$ can be interpreted as follows. To guarantee that $\Phi \succ \mu I_{KN}$, one requires extra margins in $\frac{1}{N} \sum_{i=1}^N \Psi_i$ and in the consensus gain $\alpha$, as specified in \eqref{eqn:generic_geometric_condition} and \eqref{eqn:general_consensus_gain_condition}. The extra margin $\gamma < 1$ determines how large $\alpha$ must be.

\vspace{-0.1cm}
The following lemma extends the previous result to systems modelled by a chain of integrators with a bounded input.

\vspace{-0.2cm}
\begin{lemma}
\label{lemma:lemma_distributed_consensus_observer_ISS}
Consider the system described by \eqref{eqn:general_system_dynamics_for_observer} with a bounded and piecewise continuous input $\vect{u} \in \mathbb{R}^{K}$, such that $\dot{\vect{x}}^{(M-1)} = \vect{u}$. By direct application of \autoref{lemma:ISS_stability_of_nonlinear_system}, the distributed observer given by \eqref{eqn:distributed_observer} and \eqref{eqn:distributed_observer_correction_term}, under Assumptions \ref{assumption:undirected_communication_network} and \ref{assumption:agent_measurement_model}, is ISS with respect to $\vect{u}$ provided that conditions (i), (ii) and (iii) from \autoref{theorem:theorem_distributed_consensus_observer} are satisfied.
\end{lemma}
\textbf{PROOF.} The proof is presented in \autoref{appendix:proof_lemma_distributed_consensus_observer_ISS}.

%% file: Sections/5_DistributedTargetTracking.tex
\section{Distributed target state estimation}
\label{section:distributed_target_tracking}
This section details how the proposed observer design can be used to solve the problem of bearing-based distributed target state estimation. To that end, it starts by adapting the nonlinear target measurement model introduced earlier into a more tractable model which fits the proposed estimation framework, allowing for the systematic synthesis of observers for targets with motion models characterized by an arbitrary number of integrators. This is followed by a formal analysis of high-applicability examples.

\subsection{Distributed target state observer design}
Recall the bearing measurement model introduced in \eqref{eqn:bearing_measurement}, which is restated here for the reader's convenience
\begin{equation*}
    \vect{b}_{i} = \frac{\vect{p}_\mathrm{T} -  \vect{p}_i}{\|\vect{p}_\mathrm{T} -  \vect{p}_i\|}.
\end{equation*}
This can be regarded as a nonlinear function of the system state $\vect{x}_{\mathrm{T}}$. Following \cite{BATISTA20131065}, this nonlinear model can be transformed by resorting to the orthogonal projection matrix given by
\begin{equation*}
    \vectg{\Pi}_{\vect{b}_i} \coloneqq \vect{I}_3 - \vect{b}_i \vect{b}_i^{\top},
\end{equation*}
which projects any vector $\vect{z} \in \mathbb{R}^{3}$ onto the plane orthogonal to the bearing vector $\vect{b}_i$, and satisfies
\begin{equation}
    \vectg{\Pi}_{\vect{b}_i} \left(\vect{p}_\mathrm{T} -  \vect{p}_i \right) = \vect{0}.
    \label{eqn:linear_equation_measurement}
\end{equation}
In order to define a new measurement equation as a function of the target position state, consistent with the model introduced in \eqref{eqn:general_time_varying_measurement_model}, consider the new measurement model of each agent $i$ given by
\begin{equation}
    \vect{y}_i \coloneqq \vectg{\Pi}_{\vect{b}_i}\vect{p}_{i}, \label{eqn:time_varying_measurement_model_concretization}
\end{equation}
which, according to \eqref{eqn:linear_equation_measurement}, can also be written as
\begin{equation}
    \vect{y}_i = \vectg{\Psi}_i(t, \vect{x}_{\mathrm{T}})\vect{p}_{\mathrm{T}},
    \label{eqn:time_varying_measurement_model_concretization2}
\end{equation}
with $\vectg{\Psi}_i(t, \vect{x}_{\mathrm{T}}) = \vectg{\Pi}_{\vect{b}_i}$. It follows directly that
\begin{equation}
    \vectg{\Psi}(t, \vect{x}_{\mathrm{T}}) \coloneqq \vectg{\Pi} = \diag(\vectg{\Pi}_{\vect{b}_1}, \hdots, \vectg{\Pi}_{\vect{b}_N}).
    \label{eqn:block_diagonal_projection_matrix}
\end{equation}
Define $\hat{\vect{x}}_{\mathrm{T}_i}^{(m)} \in \mathbb{R}^{3}$ as the estimate of the $m$-th component of the target state, computed at agent $i$. For notational convenience, consider also $\hat{\vect{p}}_{\mathrm{T}_i} = \hat{\vect{x}}_{\mathrm{T}_i}^{(0)}$. By direct application of the design introduced in \eqref{eqn:distributed_observer} and \eqref{eqn:distributed_observer_correction_term}, the observer dynamics are described by
\begin{equation}
	\dot{\hat{\vect{x}}}_{\mathrm{T}_i}^{(m)} =
	\begin{cases}
		\hat{\vect{x}}_{\mathrm{T}_i}^{(m+1)} + k_{m+1} \vectg{\delta}_i &\text{, if } m < M-1\\
    	k_{M} \vectg{\delta}_i &\text{, if } m=M-1
	\end{cases}
	\label{eqn:distributed_target_tracking_observer}
\end{equation}
with correction term
\begin{equation}
    \vectg{\delta}_i = \vectg{\Pi}_{\vect{b}_i}(\vect{p}_i - \hat{\vect{p}}_{\mathrm{T}_i}) - \alpha \sum_{j \in \mathcal{N}_i} a_{ij}(\hat{\vect{p}}_{\mathrm{T}_i} - \hat{\vect{p}}_{\mathrm{T}_j}),
    \label{eqn:target_observer_correction_term}
    \vspace{-0.2cm}
\end{equation}
where only the estimated target position is shared across neighbouring agents.

By replacing the measurement model \eqref{eqn:time_varying_measurement_model_concretization} in \eqref{eqn:generic_geometric_condition}, the first condition imposed by \autoref{theorem:theorem_distributed_consensus_observer} becomes equivalent to the spatial excitation condition introduced by \cite{2024_ACC}, given by
\vspace{-0.2cm}
\begin{equation}
    \frac{1}{N} \sum_{i=1}^{N} \vectg{\Pi}_{\vect{b}_i} \succ (\mu + \gamma) \vect{I}_K.
    \label{eqn:geometric_condition}
    \vspace{-0.2cm}
\end{equation}
This condition imposes a requirement on the average of the agents' projection matrices, which depend directly on the relative bearing vectors between the agents and the target. Since each matrix $\vectg{\Pi}_{\vect{b}_i}$ has a null eigenvalue associated with its corresponding bearing vector, this condition quantifies the minimal amount of global information required along each direction to guarantee the exponential convergence of the observer. In other words, it imposes restrictions on the geometric configuration of the $N$ agents and the target, and encodes the idea that a wider angular separation between agents improves the conditioning of the estimation problem. Check the work by \cite{2024_ACC} for application examples of this condition in 2D space.

By replacing \eqref{eqn:block_diagonal_projection_matrix} in \eqref{eqn:general_consensus_gain_condition} and noting that the orthogonal projection matrix $\vectg{\Pi}_{\vect{b}_i}$ is idempotent, such that the block-diagonal matrix also verifies $\vectg{\Pi}^2 = \vectg{\Pi}$, the second condition imposed by \autoref{theorem:theorem_distributed_consensus_observer} can also be simplified, leading to
\begin{equation*}
    \alpha > \frac{\mu + \left(\frac{1}{\gamma} - 1 \right)\|\vectg{\Pi}\|}{\lambda_{\min}(\vectg{\Lambda})}.
\end{equation*}
Since $\|\vectg{\Pi}\| = 1$, we can conclude that the condition of the consensus gain is given by
\begin{equation}
    \alpha > \frac{\mu + \frac{1}{\gamma} - 1}{\lambda_{\min}(\vectg{\Lambda})}.
    \label{eqn:consensus_gain_condition}
\end{equation}
Finally, the application of \autoref{theorem:theorem_distributed_consensus_observer} and \autoref{lemma:lemma_distributed_consensus_observer_ISS} with the simplifications introduced in \eqref{eqn:geometric_condition} and \eqref{eqn:consensus_gain_condition} yields the following.

\vspace{0.5cm}
\begin{corollary}
\label{corollary:target_state_observer}
Consider a target with motion model described by \eqref{eqn:target_dynamics}, and a group of $N$ agents operating under Assumptions \ref{assumption:undirected_communication_network} and \ref{assumption:agent_measurement_model} with measurement model described by \eqref{eqn:time_varying_measurement_model_concretization} and \eqref{eqn:time_varying_measurement_model_concretization2}. The origin of the estimation error associated with the distributed observer given by \eqref{eqn:distributed_target_tracking_observer} and \eqref{eqn:target_observer_correction_term} is UGES provided that inequalities \eqref{eqn:geometric_condition} and \eqref{eqn:consensus_gain_condition} are satisfied for all $t \geq t_0$, and for $M \geq 3$ the observer gains are chosen such that the LMI described by \eqref{eqn:LMI_for_theorem_1} and \eqref{eqn:gains_ratio_definition_theorem} is also satisfied. Furthermore, if the system described by \eqref{eqn:target_dynamics} has a bounded and piecewise continuous input, such that $\dot{\vect{p}}_{\mathrm{T}}^{(M-1)} = \vect{u} \in \mathbb{R}^{3}$, it follows from direct application of \autoref{lemma:lemma_distributed_consensus_observer_ISS} that the system is ISS with respect $\vect{u}$.
\end{corollary}

\subsection{Robustness to measurement loss}
Critical scenarios may arise when some agents are unable to obtain system measurements, either for brief periods of time or permanently. In such scenarios, consider that there are $L \in [0,N)$ agents that lack access to bearing measurements, for all $t \geq t_0$. As noted in \autoref{remark:no_measurements_available}, this scenario can be handled by setting the innovation term for those agents to zero, i.e., $(\vectg{\Psi}_i, \vect{y}_i)=(\vect{0},\vect{0})$, and considering a correction term given solely by the consensus term, according to 
\begin{equation}
	\vectg{\delta}_i = - \alpha \sum_{j \in \mathcal{N}_i} a_{ij} (\hat{\vect{p}}_{\mathrm{T}_i} - \hat{\vect{p}}_{\mathrm{T}_j}).
\end{equation}
By direct application of \autoref{corollary:target_state_observer}, only the spatial excitation condition imposed by \eqref{eqn:geometric_condition} becomes stricter, according to
\begin{equation}
	\frac{1}{N}\sum_{i=1}^{N-L} \vectg{\Pi}_{\vect{b}_i} \succ (\mu + \gamma)\vect{I}_K.
    \vspace{-0.2cm}
\end{equation}
This condition ensures that the collective information provided by the $N-L$ agents meets the same minimum threshold as if all $N$ agents had access to bearing measurements. The condition imposed by \eqref{eqn:consensus_gain_condition} on the consensus gain remains unchanged as its bound depends only on the maximum spectral norm of each measurement matrix. Similarly, the LMI condition described by \eqref{eqn:LMI_for_theorem_1} and \eqref{eqn:gains_ratio_definition_theorem} for $M\geq 3$ also remains unchanged, as it only depends on the proper choice of the observer gains.

\subsection{Practical examples}
\label{section:practical_examples}
\vspace{-0.1cm}
This section considers the distributed target state observer with $M=1$, $M=2$ and $M=3$. These configurations are particularly relevant in practical applications, and each presented scenario highlights the successive appearance of increasingly restrictive stability conditions that must be satisfied as the order of the target motion model increases. For the sake of simplicity, in these examples it is assumed that all agents can measure bearing vectors to the target, i.e., $L=0$.

\subsubsection{Target with constant position model ($M=1$)}
Consider a static target, with its state given by $\vect{x}_{\mathrm{T}} \coloneqq \vect{p}_{\mathrm{T}} \in \mathbb{R}^3$ such that $\dot{\vect{p}}_{\mathrm{T}} = \vect{0}$. It follows directly from \autoref{corollary:target_state_observer} that the proposed distributed observer is given by
\vspace{-0.2cm}
\begin{equation*}
    \dot{\hat{\vect{p}}}_{\mathrm{T}_i} = k_1 \left(\vectg{\Pi}_{\vect{b}_i}(\vect{p}_i - \hat{\vect{p}}_{\mathrm{T}_i}) - \alpha \sum_{j \in \mathcal{N}_i} a_{ij}(\hat{\vect{p}}_{\mathrm{T}_i} - \hat{\vect{p}}_{\mathrm{T}_j})\right).
\end{equation*}
This particular observer corresponds to the case covered in \cite{2024_ACC}, and the exponential stability results hold for any positive gain $k_1$ if there exists $\delta, \gamma > 0$ for all $t \geq t_0$ such that the geometric condition
\begin{equation*}
    \frac{1}{N} \sum_{i=1}^{N} \vectg{\Pi}_{\vect{b}_i} \succ (\delta + \gamma) \vect{I}_3,
\end{equation*}
is satisfied. This result follows from \autoref{theorem:theorem_distributed_consensus_observer} which states that for $M=1$, $\mu=\delta$. It also follows from the proof of \autoref{theorem:stability_of_nonlinear_system}, in \autoref{appendix:proof_theorem_stability_of_nonlinear_system}, that for $M=1$ the exponential convergence rate of the observer is only dictated by $\delta k_1$, according to \eqref{eqn:rate_of_convergence_M1}. Increasing the gain $k_1$ will accelerate the exponential convergence of the observer, but it will also amplify the effects of noise on the bearing measurements. Additionally, by strategically distributing the agents around the target to maximize the minimum angle between bearing vectors, it is possible to increase the minimum admissible value for $\delta$. To illustrate this concept, consider the particular case of $N=2$. In this scenario, the spatial excitation condition translates directly into ensuring that the minimal angle formed between the bearing vectors of the two agents is within a threshold dictated by $\delta$ and $\gamma$.

Additionally, the condition on the consensus gain
\vspace{-0.1cm}
\begin{equation*}
    \alpha > \frac{\delta + \frac{1}{\gamma} -1}{\lambda_{\min}(\vectg{\Lambda})},
    \vspace{-0.1cm}
\end{equation*}
must also be satisfied. In this example, we can see the interplay between the parameters $\delta$ and $\gamma$, and the minimal consensus gain $\alpha$ that ensures exponential stability. Notice that increasing the prescribed convergence rate dictated by $\delta$, according to \eqref{eqn:rate_of_convergence_M1}, directly imposes a higher consensus gain. In addition, a higher value of $\gamma$ imposes a smaller value on $\alpha$, but also increases the angular constraints imposed on the agents' formation. This interplay of parameters provides a blueprint for the observer gain design. Given that $\lambda_{\text{max}}(\vectg{\Pi}_{\vect{b}_i}) = 1$, it is straightforward that the sum of parameters should always satisfy $\delta + \gamma < 1$. Since $\gamma < 1$, this parameter will have the most impact on the consensus gain, and it should be designed to be as small as possible, reducing its impact on the spatial excitation requirements, while ensuring a reasonable value for $\alpha$. The gain $k_1$ should be designed to be as high as possible, while taking into account that in a real world scenario this will also amplify the bearing measurement noise.

\subsubsection{Target with constant velocity model ($M=2$)}
Consider a target with a constant velocity motion model, with state given by $\vect{x}_{\mathrm{T}}\coloneqq[\vect{p}_{\mathrm{T}}^{\top}, \vect{v}_{\mathrm{T}}^{\top}]^{\top} \in \mathbb{R}^{6}$, where we define $\vect{x}_{\mathrm{T}}^{(1)} \coloneqq \vect{v}_{\mathrm{T}}$ for notational convenience. The dynamics of the system are described by a double integrator according to $\dot{\vect{p}}_{\mathrm{T}} = \vect{v}_{\mathrm{T}}$ and $\dot{\vect{v}}_{\mathrm{T}} = \vect{0}$. Following the proposed methodology, the distributed observer is given by
\begin{equation*}
\begin{cases}
    \dot{\hat{\vect{p}}}_{\mathrm{T}_i} &= \hat{\vect{v}}_{\mathrm{T}_i} + k_1 \vectg{\delta}_i\\
    \dot{\hat{\vect{v}}}_{\mathrm{T}_i} &= k_2 \vectg{\delta}_i
\end{cases},
\end{equation*}
where the correction term $\vectg{\delta}_i(t)$ is given by \eqref{eqn:target_observer_correction_term}. In this scenario, the origin of the estimation error is UGES if there exists $\delta, \gamma > 0$ such that for all $t \geq t_0$  the geometric condition satisfies
\begin{equation}
    \frac{1}{N} \sum_{i=1}^{N} \vectg{\Pi}_{\vect{b}_i} \succ \left(\frac{\delta k_1 + k_2}{k_1^2} + \gamma \right) \vect{I}_3.
    \label{eqn:final_practical_geometrical_constraint}
\end{equation} 
Unlike in the previous example, the gains $k_1$ and $k_2$ chosen for the innovation term will also impact the spatial constrains imposed on the agents' formation relative to the target. Additionally, the consensus gain must also satisfy
\begin{equation}
    \alpha > \frac{\frac{\delta k_1 + k_2}{k_1^2} + \left( \frac{1}{\gamma} - 1\right)}{\lambda_{\min}(\vectg{\Lambda})},
    \label{eqn:final_pratical_constraint_on_consensus_gain}
\end{equation}
which is also affected by the gains $k_1$ and $k_2$. Borrowing intuition from linear control theory, we can regard $k_1$ as proportional gain and $k_2$ as a derivative gain. As such, the stability conditions hint at a choice of gains where $k_1 > k_2$, in order to weaken the restrictive geometrical conditions imposed on the formation and minimize the consensus gain. It also follows directly from the proof of \autoref{theorem:stability_of_nonlinear_system}, in \autoref{appendix:proof_theorem_stability_of_nonlinear_system}, and \autoref{remark:M2_LMI_is_satistisfied}, that for $M=2$, with positive values of $k_1, k_2$ and $\delta$, $\bar{\vect{Q}} \succ 0$ and the rate of exponential decay is bounded by $\text{min}(\delta, k_2/k_1)$, according to \eqref{eqn:rate_of_convergence_M2}. This in turn implies that reducing $ k_2/k_1$ below the designed value for $\delta$ will have a negative impact on the prescribed exponential decay of the observer error.

\subsubsection{Target with constant acceleration model ($M=3$)}
Similar to the previous scenarios, consider a target following a constant acceleration motion model, with state given by $\vect{x}_{\mathrm{T}}\coloneqq[\vect{p}_{\mathrm{T}}^{\top}, \vect{v}_{\mathrm{T}}^{\top}, \vect{a}_{\mathrm{T}}^{\top}]^{\top} \in \mathbb{R}^{9}$, with dynamics described by $\dot{\vect{p}}_{\mathrm{T}} = \vect{v}_{\mathrm{T}}$, $\dot{\vect{v}}_{\mathrm{T}} = \vect{a}_{\mathrm{T}}$ and $\dot{\vect{a}}_{\mathrm{T}} = \vect{0}$. The proposed observer is given by
\begin{equation*}
\begin{cases}
    \dot{\hat{\vect{p}}}_{\mathrm{T}_i} &= \hat{\vect{v}}_{\mathrm{T}_i} + k_1 \vectg{\delta}_i\\
    \dot{\hat{\vect{v}}}_{\mathrm{T}_i} &= \hat{\vect{a}}_{\mathrm{T}_i} + k_2 \vectg{\delta}_i\\
    \dot{\hat{\vect{a}}}_{\mathrm{T}_i} &= k_3 \vectg{\delta}_i\\
\end{cases},
\end{equation*}
with $\vectg{\delta}_i$ given by \eqref{eqn:target_observer_correction_term}. In this scenario, and for any case where $M \geq 3$, the origin of the estimation error is UGES if the geometrical condition imposed by \eqref{eqn:final_practical_geometrical_constraint} is satisfied, and the choice of the consensus gain $\alpha$ also satisfies \eqref{eqn:final_pratical_constraint_on_consensus_gain}. Unlike the previous examples, the addition of an extra state derivative does not change the geometric and consensus conditions imposed on the system. Instead, it follows from \autoref{theorem:theorem_distributed_consensus_observer} and \autoref{remark:M2_LMI_is_satistisfied} that for $M \geq 3$ only an additional LMI constraint that depends on the ratio of system gains and the parameter $\delta$ is imposed. For this particular example, this inequality is given by
\begin{equation}
    \bar{\vect{Q}}
    = 2
    \begin{bmatrix}
        \frac{k_3}{k_2} & 0 & \frac{k_3}{2k_2} \\
        0 & \frac{k_2}{k_1} - \frac{k_3}{k_2} & -\frac{k_3}{2k_2} \\
        \frac{k_3}{2k_2} & -\frac{k_3}{2k_2} & \delta
    \end{bmatrix} 
    \succ 0.
    \label{eqn:LMI_for_constant_acceleration_example}
\end{equation} 
This structure provides insights for choosing the gains of the system such that they satisfy $k_1 > k_2 > k_3$, for which higher values of the ratio $k_2/k_1$ will leave a higher design margin for the gain $k_3$, with the trade-off of also increasing the required angular separation between the formation of agents imposed by \eqref{eqn:final_practical_geometrical_constraint}. 

The addition of extra derivatives to the target motion model also impacts the prescribed rate of convergence of the system, as it follows directly from the proof of \autoref{theorem:stability_of_nonlinear_system}, in \autoref{appendix:proof_theorem_stability_of_nonlinear_system} that for $M\geq 2$, the prescribed bound on the convergence rate is dictated by $\lambda_{\min}(\bar{\vect{Q}})$.

\subsection{Parameter and gain selection for $M \geq 2$}
To choose the system gains, we can start by selecting a set of values for the parameters $\delta$ and $\gamma$. The value of $\delta$ will have the most impact in the spatial excitation condition and the convergence rate of the observer. In practice it should be small, i.e. $\delta < 1$, and find a balance between penalizing the convergence rate or the geometric condition imposed by \eqref{eqn:final_practical_geometrical_constraint}. The value of $\gamma$ will have the most impact on the consensus gain, according to \eqref{eqn:final_pratical_constraint_on_consensus_gain} -- it should also be small (e.g. $\gamma < 0.5$), to reduce the impact on the geometric condition, while ensuring a reasonable minimum value for $\alpha$. Then, we can proceed to choose the gain $k_1$ and the ratio $k_2/k_1$, such that the remaining ratios of gains $k_{i+1}/k_i, i=2,\hdots,M-1$ satisfy the LMI given by \eqref{eqn:LMI_for_theorem_1} and \eqref{eqn:gains_ratio_definition_theorem}. This can be achieved by resorting to solvers such as YALMIP (see \cite{yalmip}). Finally, the value for the consensus gain $\alpha$ can be chosen by direct application of \eqref{eqn:final_pratical_constraint_on_consensus_gain}.

%% file: Sections/6_NumericalResults.tex
\section{Numerical results}
\label{section:SimulationResults}
\begin{figure}
	\centering
    \includegraphics[width=0.30\textwidth]{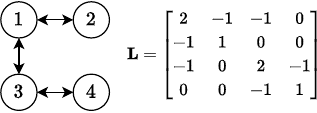}
	\caption{Graph and associated Laplacian adopted for the numerical examples.}
	\label{fig:graph_and_laplacian}
\end{figure}

This section presents simulation results to illustrate the performance of the proposed observer design when applied to some of the practical examples presented in this paper. The communication topology adopted for the results is composed of $N=4$ agents and is modelled according to \autoref{fig:graph_and_laplacian}. The agents are arranged geometrically in a square formation with $\vect{p}_{1} = [-10, \, 10, \, 2]^{\top} \SI{}{m}$, $\vect{p}_{2} = [10, \, 10, \, 2]^{\top} \SI{}{m}$, $\vect{p}_{3} = [10, \, -10, \, 2]^{\top} \SI{}{m}$ and $\vect{p}_{4} = [-10, \, -10, \, 2]^{\top} \SI{}{m}$. For simplicity, all agent have access to their own position $\vect{p}_{i}$ corrupted by Gaussian noise with a standard deviation of $0.1 \SI{}{m}$ and bearing measurements corrupted by rotation noise with an angle standard deviation of $1.0^{\circ}$.

\subsection{Target with constant velocity}
\label{section:numerical_results:target_with_constant_velocity}
Consider a scenario where a target with $\vect{p}_{\mathrm{T}}(t_0) = [0, \, -15, \, 0]^{\top} \SI{}{m}$ moves with constant velocity $\vect{v}_{\mathrm{T}} = [0, \, 0.5, \, 0]^{\top} \SI{}{m/s}$. Consider also the choice of observer gains $k_1=5.0$, $k_2=3.5$ and $\alpha=15.9$, with parameters $\gamma=0.1$ and $\delta = 0.8$, designed under the assumption that the conditions given by $\frac{1}{4}\sum_{i=1}^{4} \vectg{\Pi}_{\vect{b}_i} \succ 0.4 \vect{I}_3$, and \eqref{eqn:final_pratical_constraint_on_consensus_gain} are satisfied.

\begin{figure}
	\centering
    \includegraphics[width=0.48\textwidth]{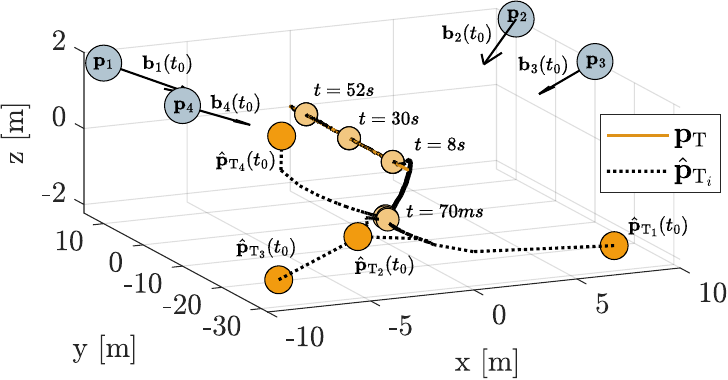}
	\caption{Three-dimensional representation of four agents (in blue circles) tracking a target moving with constant velocity. The orange markers correspond to discrete time samples of the target position estimates computed by each agent, which converge to the same value after a short period of time.}
	\label{fig:constant_velocity:3d_plot}
\end{figure}

The distributed observer is initialized by considering initial position estimates that are aligned with the initial bearing measurements, according to $\hat{\vect{p}}_{\mathrm{T}_i}(t_0) = \vect{p}_i + r_i \vect{b}_i(t_0) $, where $r_i > 0$ is an initial random range, and $\hat{\vect{v}}_{\mathrm{T}_i}(t_0) = [0, \, 0, \, 0]^{\top} \SI{}{m/s}$ for all agents. \autoref{fig:constant_velocity:3d_plot} depicts a 3D representation of the position of each agent, and the evolution of the target position estimates, with the blue markers corresponding to the four agents position, and the orange markers to discrete time samples of the position estimates provided by each agent.

\begin{figure}
    \centering
    \begin{subfigure}{0.230\textwidth}
        \centering
        \includegraphics[width=\linewidth]{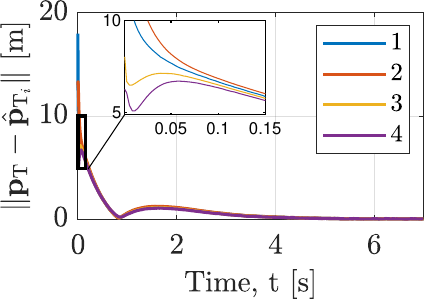}
        \caption{Position estimation error.}
        \label{subfig:constant_velocity:position_error}
    \end{subfigure}
    \hfill
    \begin{subfigure}{0.230\textwidth}
        \centering
        \includegraphics[width=\linewidth]{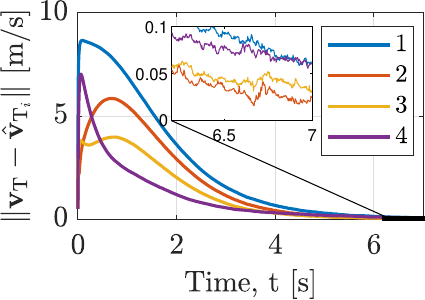}
        \caption{Velocity estimation error.}
        \label{subfig:constant_velocity:velocity_error}
    \end{subfigure}
    \caption{Evolution of the position and velocity estimation errors for each individual agent.}
    \label{fig:constant_velocity:velocity_error}
\end{figure}

In \autoref{subfig:constant_velocity:position_error}, it can be observed that the position estimates of all four agents rapidly reach consensus, and the position error converges exponentially to zero. In contrast, the velocity estimates do not reach consensus instantly and take longer to converge, according to \autoref{subfig:constant_velocity:velocity_error}. This is expected, as the correction term does not explicitly enforce consensus on the estimated velocity, and this is achieved by implicitly leveraging the chain of integrators in the system model. 

\begin{figure}
    \centering
    \begin{subfigure}{0.230\textwidth}
        \centering
        \includegraphics[width=\linewidth]{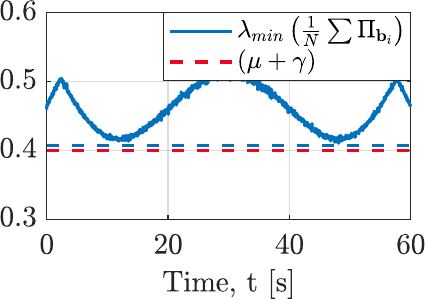}
        \caption{Evolution of the minimum eigenvalue associated with the spatial excitation condition.}
        \label{subfig:constant_velocity:minimum_eigenvalue}
    \end{subfigure}
    \hfill
    \begin{subfigure}{0.222\textwidth}
        \centering
        \includegraphics[width=\linewidth]{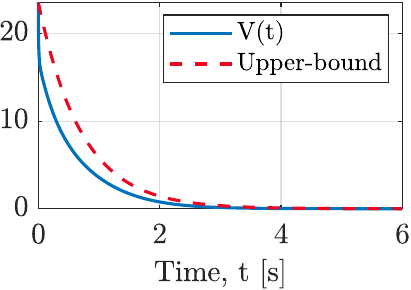}
        \caption{Decay of the Lyapunov function associated with \autoref{theorem:stability_of_nonlinear_system}, and its upper-bound.}
        \label{subfig:constant_velocity:lyapunov_bound}
    \end{subfigure}
    \caption{Variation of the spatial excitation and decay of the Lyapunov function.}
    \label{fig:lyapunov_and_conservative_bounds}
\end{figure}

The time evolution of the minimum eigenvalue associated with the spatial excitation condition is shown in \autoref{subfig:constant_velocity:minimum_eigenvalue}, with a dashed red line indicating the conservative value adopted for $\mu + \delta=0.4$. \autoref{subfig:constant_velocity:lyapunov_bound} showcases the exponential decay of the Lyapunov candidate function introduced in the stability proof of \autoref{theorem:stability_of_nonlinear_system}, along with its upper-bound.

\begin{figure}
    \centering
    \begin{subfigure}[t]{0.235\textwidth}
        \centering
        \includegraphics[width=\linewidth]{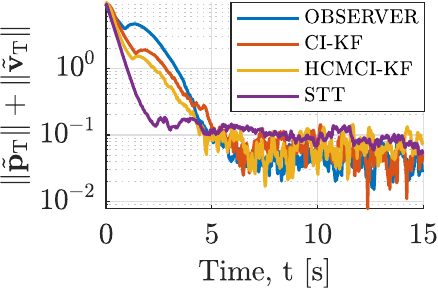}
        \caption{Evolution of the estimation error for the DKF methods, STT and the observer.}
        \label{subfig:constant_velocity:comparison_convergence}
    \end{subfigure}
    \begin{subfigure}[t]{0.235\textwidth}
        \centering
        \includegraphics[width=\linewidth]{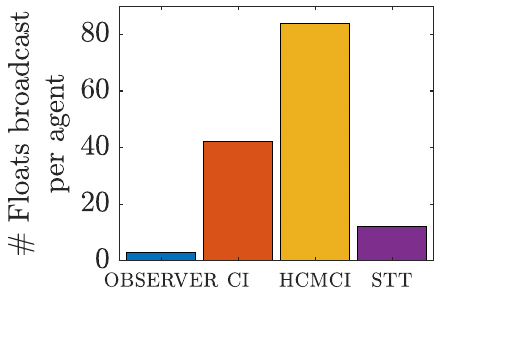}
        \caption{Number of floats broadcast to each neighbouring agent by each method.}
        \label{subfig:constant_velocity:comparison_comms}
    \end{subfigure}
    \caption{Comparison between multiple DKF methods, STT and the proposed observer (for $M=2$) of the evolution of the estimation error and the total information broadcast by each agent.}
    \label{fig:constant_velocity:comparison_DKF}
\end{figure}

The performance of the proposed observer is also compared with two variants of the DKF method (CI-KF and an HCMCI-KF), proposed in Tables II and V of \cite{battistelli_2015}. The process and measurement noise covariance matrices adopted for the DKFs are given by $\vect{Q}=\vect{I}_6$, and $\vect{R}=0.007\vect{I}_3$, respectively, and the covariance matrix is initialized according to $\vectg{\Omega}_i(t_0)=\vect{I}_6$, for all $i=1,\hdots,4$ agents. Additionally, the observer performance is also compared with the Spatial-Temporal Triangulation (STT) method proposed in \cite{ZHENG2025112216}, with $\gamma_1=7.8$, $\gamma_2 = 6.3$ and $c=2.5$. To make the comparison as fair as possible, the DKFs were tuned to perform only two consensus iterations per simulation time-step. All algorithms were initialized with the same position estimates, aligned with the initial bearing measurements. \autoref{subfig:constant_velocity:comparison_convergence} showcases that with proper initialization and gains choice, all methods have a comparable convergence rate, with the proposed observer converging slightly slower than the other methods. In contrast, the proposed observer requires less data to be shared across the network, with each agent only broadcasting a 3-dimensional vector to its neighbouring agents, corresponding to the estimated target position. An optimized implementation of the STT method requires each agent to broadcast a total of 12 floats to its neighbours, corresponding to the 3-dimensional bearing measurement, a 6-dimensional estimated target state and 3-dimensional position of the agent. The CI-KF method requires each agent to share an information pair composed of a matrix with 36 entries, and a 6-dimensional vector, for each consensus iteration and the HCMCI-KF method requires twice as much information to be broadcast to the network, according to \autoref{subfig:constant_velocity:comparison_comms}. This example demonstrates that the proposed observer exhibits a good trade-off between communication bandwidth, exponential convergence properties and explicit stability conditions.

\begin{figure}
    \centering
    \begin{subfigure}{0.230\textwidth}
        \centering
        \includegraphics[width=\linewidth]{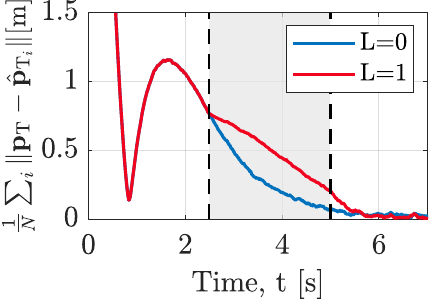}
        \caption{Average position error.}
        \label{subfig:constant_velocity:position_multi_L}
    \end{subfigure}
    \hfill
    \begin{subfigure}{0.230\textwidth}
        \centering
        \includegraphics[width=\linewidth]{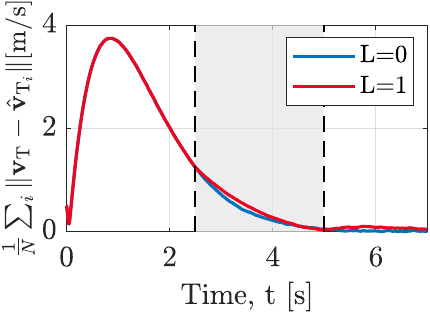}
        \caption{Average velocity error.}
        \label{subfig:constant_velocity:velocity_multi_L}
    \end{subfigure}
    \caption{Comparison of the average position and velocity estimation errors when one agent suffers a temporary loss of measurements, i.e. $L=1$.}
    \label{fig:constant_velocity_multi_l}
\end{figure}

To illustrate the robustness of the observer to measurement losses, the previous experiment is repeated, but with one agent unable to obtain system measurements, i.e. $L=1$, for $t \in [2.5, 5.0] \SI{}{s}$. In \autoref{fig:constant_velocity_multi_l}, it can be observed that the average position and velocity errors continue to decrease during this time-interval, although at a slower rate when compared to the original scenario where no loss of measurements is considered. Since the gain $k_2 < k_1$, the impact on the rate of convergence will be greater for the position error, when compared to the velocity error.

\begin{figure}
    \centering
    \begin{subfigure}[t]{0.230\textwidth}
        \centering
        \includegraphics[width=\linewidth]{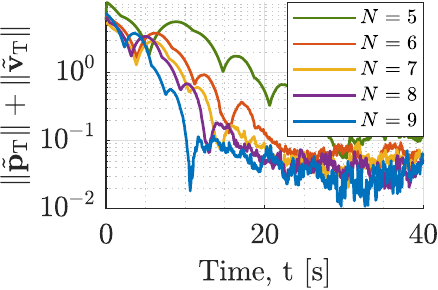}
        \caption{Agents arranged in a circle, separated by their nearest neighbour by $20^{\circ}$.}
        \label{subfig:more_than_four_agents:good_agent_distribution}
    \end{subfigure}
    \hfill
    \begin{subfigure}[t]{0.230\textwidth}
        \centering
        \includegraphics[width=\linewidth]{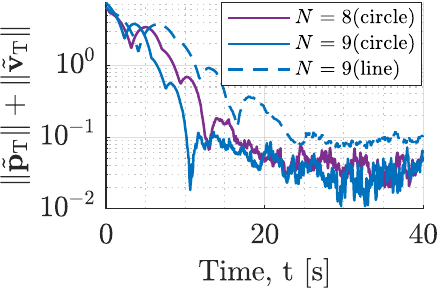}
        \caption{Agents arranged in a line, compared to the circle arrangement on the left.}
        \label{subfig:more_than_four_agents:bad_agent_distribution}
    \end{subfigure}
    \caption{Observer's performance with an increasing number of agents and different geometric distributions relative to the target.}
    \label{fig:more_than_four_agents}
\end{figure}

To highlight the impact of the number of agents $N$ and their spatial distribution, consider a similar scenario with the constant velocity target, but with $N$ agents positioned on a horizontal circle with a $10 \SI{}{m}$ radius, centered at the origin, at a height of $2 \SI{}{m}$. Agents are spaced $20^{\circ}$ apart from their nearest neighbour and only exchange data with its two nearest neighbours. As shown in \autoref{subfig:more_than_four_agents:good_agent_distribution}, increasing the number of agents in this circular arrangement improves the observer convergence rate. In contrast, \autoref{subfig:more_than_four_agents:bad_agent_distribution} demonstrates that a linear arrangement, i.e. $N=9$ agents aligned along the target's trajectory, each separated by $2\SI{}{m}$, can yield a worse convergence performance compared to $N=8$ in a circular configuration. This demonstrates the importance of a proper agent distribution to achieving a faster convergence rate, in accordance with \autoref{theorem:theorem_distributed_consensus_observer}.

\subsection{Target with constant acceleration}
In this section, we demonstrate the performance of the proposed observer with $M=3$, and compare it to a mismatched model of the target with $M=2$, illustrating the system ISS properties. Consider a new scenario, where the target with $\vect{p}_{\mathrm{T}}(t_0) = [0, \, 10, \, 0]^{\top} \SI{}{m}$ and $\vect{v}_{\mathrm{T}}(t_0) = [0, \, -2, \, 0]^{\top} \SI{}{m/s}$ moves with constant acceleration $\vect{a}_{\mathrm{T}} = [0, \, 0.15, \, 0.01]^{\top} \SI{}{m/s^2}$. 

\subsubsection{Distributed observer with $M=3$}
\begin{figure*}
    \centering
    \begin{subfigure}{0.275\textwidth}
        \centering
        \includegraphics[width=0.860\linewidth]{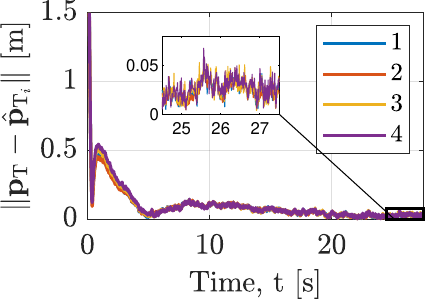}
        \caption{Position estimation error.}
        \label{subfig:constant_acceleration:M3:position_error}
    \end{subfigure}
    \hspace{0.13cm}
    \begin{subfigure}{0.275\textwidth}
        \centering
        \includegraphics[width=0.860\linewidth]{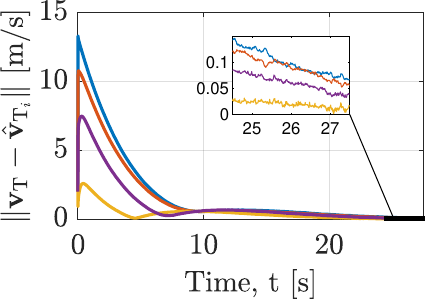}
        \caption{Velocity estimation error.}
        \label{subfig:constant_acceleration:M3:velocity_error}
    \end{subfigure}
    \hspace{0.13cm}
    \begin{subfigure}{0.275\textwidth}
        \centering
        \includegraphics[width=0.860\linewidth]{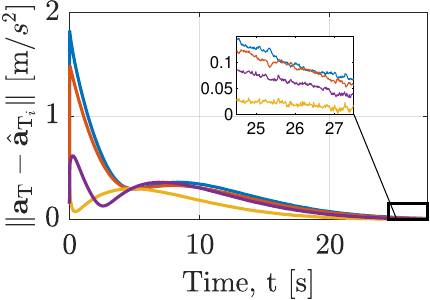}
        \caption{Acceleration estimation error.}
        \label{subfig:constant_acceleration:M3:acceleration_error}
    \end{subfigure}
    \vspace{-0.2cm}
    \caption{Example of the convergence properties of the observer, for a target moving with constant acceleration and $M=3$.}
    \label{fig:constant_acceleration:M3}
\end{figure*}

Similar to the first example, the initial observer estimates for the target velocity and acceleration are initialized to zero, and the position estimates are aligned with the initial bearing measurements. In this example, the distributed observer gains $k_1=10$, $k_2=3.7$, $k_3=0.5$ and $\alpha=15.5$ were designed assuming $\gamma = 0.1$ and $\delta=0.3$, such that the LMI given by \eqref{eqn:LMI_for_constant_acceleration_example} is satisfied. The exponential decay of the position, velocity and acceleration errors for each agent is illustrated in \autoref{fig:constant_acceleration:M3}. Despite the higher gain $k_1$ the observer takes longer to converge than in the previous examples. This demonstrates the trade-off associated with increasing the model order: while it can improve the estimation accuracy, it also increases the convergence time due to the addition of an extra integrator in the system, which is turn is associated with the addition of a smaller gain $k_3$. Increasing the observer gains can improve the convergence rate, but at the cost of amplifying the noise in the bearing measurements.

\subsubsection{Distributed observer with $M=2$}
\begin{figure*}
    \centering
    \begin{subfigure}{0.275\textwidth}
        \centering
        \includegraphics[width=0.860\linewidth]{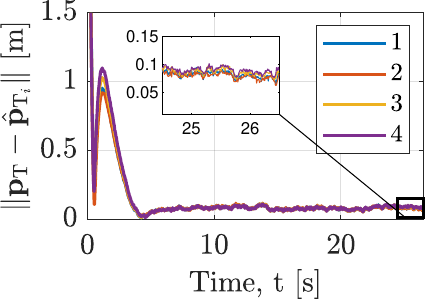}
        \caption{Position estimation error.}
        \label{subfig:constant_acceleration:M2:position_error}
    \end{subfigure}
    \hspace{0.13cm}
    \begin{subfigure}{0.275\textwidth}
        \centering
        \includegraphics[width=0.860\linewidth]{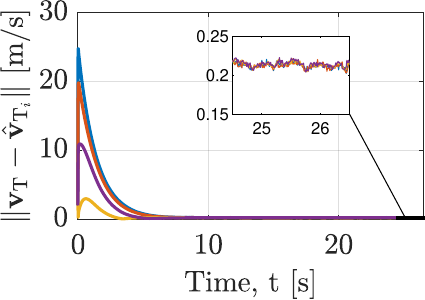}
        \caption{Velocity estimation error.}
        \label{subfig:constant_acceleration:M2:velocity_error}
    \end{subfigure}
    \vspace{-0.2cm}
    \caption{Example of the ISS properties of the observer, for a target moving with constant acceleration and $M=2$.}
    \label{fig:constant_acceleration:M2}
\end{figure*}

To demonstrate the ISS properties of the proposed method, consider the observer with $M=2$, introduced in \autoref{section:numerical_results:target_with_constant_velocity}, used to estimate the position and velocity of a target moving with constant acceleration. \autoref{fig:constant_acceleration:M2} demonstrates that the position and velocity errors are bounded and converge to a ball around the origin, dependent on the target acceleration, which can be interpreted as the bounded input $\vect{u}\in \mathbb{R}^{3}$ of the system, when $M=2$. While the observer errors remain bounded, the convergence rate is slightly faster for $M=2$ than for $M=3$, highlighting that increasing the observer order to better match the target motion reduces the estimation error, but also decreases the convergence rate. This trade-off should be taken into consideration during the design phase, when selecting the observer order $M$.

\subsection{Higher-order observer models}
\label{section:numerical_results:higher_order_observer_models}
\begin{figure}
    \centering
    \begin{subfigure}[t]{0.23\textwidth}
        \centering
        \includegraphics[width=\linewidth]{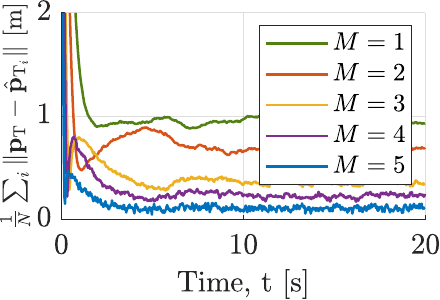}
        \caption{Average position error.}
        \label{subfig:circle_trajectory:position_error}
    \end{subfigure}
    \begin{subfigure}[t]{0.23\textwidth}
        \centering
        \includegraphics[width=\linewidth]{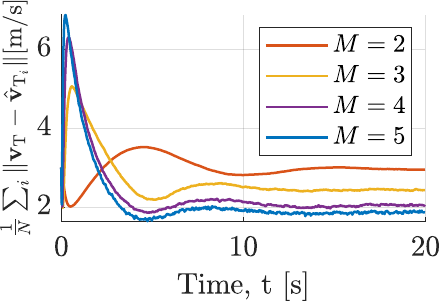}
        \caption{Average velocity error.}
        \label{subfig:circle_trajectory:velocity_error}
    \end{subfigure}
    \begin{subfigure}[t]{0.23\textwidth}
        \centering
        \includegraphics[width=\linewidth]{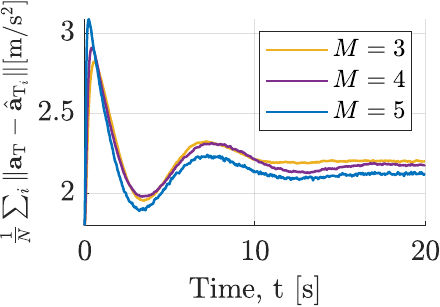}
        \caption{Average acceleration error.}
        \label{subfig:circle_trajectory:acceleration_error}
    \end{subfigure}
    \begin{subfigure}[t]{0.23\textwidth}
        \centering
        \includegraphics[width=\linewidth]{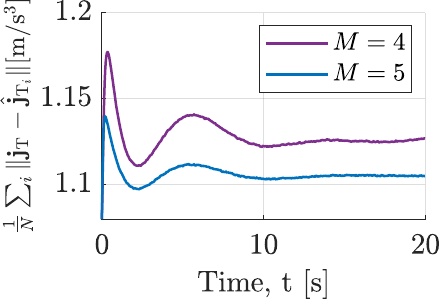}
        \caption{Average jerk error.}
        \label{subfig:circle_trajectory:jerk_error}
    \end{subfigure}
    \caption{Average observer estimation errors, for a target moving in a circular trajectory at a speed of 3 m/s.}
    \label{fig:circle_trajectory}
\end{figure}

In this section we analyse the performance of the observer when considering higher-order motion models. Consider the agents arranged in a square formation, as previously described, and a target that executes a circular motion given by $\vect{p}_{\mathrm{T}}(t) = [5\cos(0.6t), \, 5\sin(0.6t), \, 0]^{\top}$. The observer gains where chosen for each observer such that $\mu + \gamma \approx 0.4$ and $\alpha \approx 2.6$. In \autoref{fig:circle_trajectory}, we can observe the evolution of the average position, velocity, acceleration and jerk estimation errors for $M = 1,\hdots,5$. Notice that in this example, higher-order time derivatives of the target position are not zero and none of motion models adopted matches exactly the motion executed by the target, further highlighting the ISS properties of the system. Observe also that with a higher order model, we are able to better estimate the real motion of the target, reducing the bounded estimation error. However, increasing the model order also increases the noise standard deviation in the position estimate and its subsequent time derivatives.

\subsection{Robustness to measurement noise}
\begin{figure}
    \centering
    \begin{subfigure}[t]{0.230\textwidth}
        \centering
        \includegraphics[width=\linewidth]{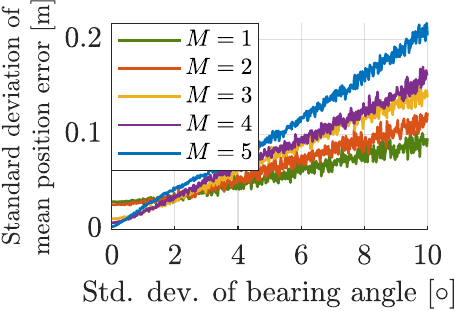}
        \caption{Standard deviation of the average position estimation error for four agents.}
        \label{subfig:circle_trajectory:error_standard_deviation}
    \end{subfigure}
    \hfill
    \begin{subfigure}[t]{0.230\textwidth}
        \centering
        \includegraphics[width=\linewidth]{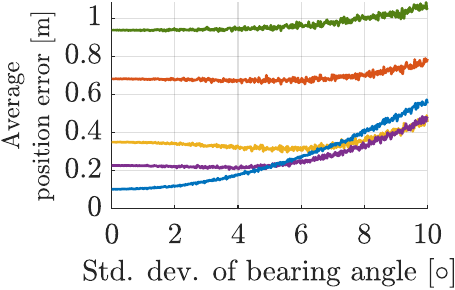}
        \caption{Average position estimation error for four agents.}
        \label{subfig:circle_trajectory:error_mean_deviation}
    \end{subfigure}
    \caption{Standard deviation and mean of the average position estimation errors for four agents as a function of the standard deviation of the noise in the bearing measurement, for multiple motion model orders.}
    \label{fig:circle_trajectory:error_noise}
\end{figure}

To further investigate the robustness of the observer to measurement noise, the experiment described in the previous section was run for multiple levels of noise in the bearing vector. In \autoref{fig:circle_trajectory:error_noise}, we present the standard deviation and average position errors as functions of the bearing angle noise standard deviation. In \autoref{subfig:circle_trajectory:error_standard_deviation}, it can be observed that higher-order models exhibit a greater line slope, which translates directly to higher noise amplification. From \autoref{subfig:circle_trajectory:error_mean_deviation}, it can also be observed that for low measurement noise, lower-order models exhibit a higher average position estimation error, consistent with previous results. Notwithstanding, after a given noise threshold, there is a trade-off between increasing the order of the observer and amplifying the measurement noise. For example, if the measurement standard deviation of the bearing is greater than $5.5^{\circ}$, then the average position error will be higher with $M=5$ than with $M=4$.

%% file: Sections/7_Conclusion.tex
\section{Conclusion}
\label{section:Conclusion}
This work introduced a novel observer design to solve the problem of distributed target state estimation with bearing measurements. It started by introducing a general fixed-gain consensus-based observer design for estimating the state of systems modelled by a chain of integrators of arbitrary order, with a particular class of nonlinear measurement models in an undirected communication network of agents. The proposed solution leveraged the integrator structure of the system dynamics in order to reduce the data exchanged in the network, requiring each agent to broadcast only a component of the state estimate. By exploiting the properties of the orthogonal projection matrix, the proposed method was then used to systematically solve the distributed target tracking problem for a target with an arbitrary motion model order. Explicit conditions on the system gains and geometric conditions on the bearing measurements were provided for obtaining exponential stability properties. Robustness to measurement loss was addressed, and practical examples where derived for a target modelled with first-, second-, and third-order integrator dynamics, highlighting the systematic design procedure and the growing strictness of the stability conditions imposed. To conclude, simulation results were presented to illustrate the performance of the proposed algorithm. Future work includes further research into the stability conditions and convergence properties of the proposed observer under scenarios where the derived geometric conditions for the agents' formation are not satisfied, but there is enough relative motion between the agents and the target to resort to PE arguments. Future efforts also include the coupling of the proposed observer with a cooperative motion controller. This is crucial to guarantee the observability of the system under challenging environments where occlusions might occur and ensure that the geometrical conditions are always satisfied by the agents.

%% file: Sections/Appendix/SchurLemma.tex
\section{Proof of \autoref{lemma:schur_inverse_trick}}
\label{appendix:proof_lemma_schur_inverse_trick}
\textbf{Proof.} Let $\vect{C} \succ \gamma \vect{I}$, from which it follows directly that $\frac{1}{\gamma}\vect{I} - \vect{C}^{-1} \succ 0$. If we multiply the matrix inequality by $\vect{B}$ and $\vect{B}^{\top}$ on the left and right, respectively, we get $\frac{1}{\gamma}\vect{B}\vect{B}^\top \succeq \vect{B}\vect{C}^{-1}\vect{B}^\top$, from which we directly conclude that $\vect{A} - \frac{1}{\gamma}\vect{B}\vect{B}^{\top} \succ 0 \implies \vect{A} - \vect{B}\vect{C}^{-1}\vect{B}^{\top} \succ 0$. \qed

%% file: Sections/Appendix/StabilityAnalysis.tex
\section{Proof of \autoref{theorem:stability_of_nonlinear_system}}
\label{appendix:proof_theorem_stability_of_nonlinear_system}
\vspace{-0.2cm}
\textbf{Proof.} Observe that the block $\vectg{\Phi}(t,\vectg{\xi})$ appears repeatedly along the first column of the system dynamics matrix in \eqref{eqn:time_variant_matrix_error_structure}. To isolate this term in a single matrix block, consider a time-invariant similarity transformation that preserves the properties of the system, given by 
\begin{equation}
    \vectg{\eta} = \vect{P} \vectg{\xi},
    \label{eqn:similarity_transformation}
\end{equation}
where $\vectg{\eta} \coloneqq [\vectg{\eta}^{(0)}, \hdots, \vectg{\eta}^{(M-1)}]^{\top} \in \mathbb{R}^{WM}$ and 
\begin{equation*}
    \vect{P} \coloneqq 
    \begin{bmatrix}
        \vect{0} & \vect{0} & \hdots & -\frac{1}{k_{M-1}}\vect{I}_W &  \frac{1}{k_M}\vect{I}_W\\
        \vect{0} & \vect{0} & \hdots & \frac{1}{k_{M-1}}\vect{I}_W & \vect{0}\\
        \vdots   & \vdots & \ddots & \vdots & \vdots \\ 
        \vect{0} & -\frac{1}{k_2}\vect{I}_W & \hdots & \vect{0} & \vect{0}\\
        -\frac{1}{k_1}\vect{I}_W & \frac{1}{k_2}\vect{I}_W & \hdots & \vect{0} & \vect{0}\\
        \frac{1}{k_1}\vect{I}_W & \vect{0} & \hdots & \vect{0} & \vect{0}
    \end{bmatrix}.
    \label{eqn:state_transformation_matrix_P}
\end{equation*}
Consider the new system dynamics dictated by $\dot{\vectg{\eta}} = \vectg{\Sigma}(t, \vectg{\eta})\vectg{\eta}$, where
\begin{equation}
    \vectg{\Sigma}(t, \vectg{\eta}) \coloneqq \vect{P} \vectg{\Xi}(t, \vectg{\xi}) \vect{P}^{-1} \text{ and } \vectg{\xi} = \vect{P}^{-1} \vectg{\eta}.
\end{equation}
From this definition, it follows directly that the matrix $\vectg{\Sigma}(t, \vectg{\eta})$ can also be expressed according to the following assertions:
\begin{enumerate}[label=(\roman*)]
    \item for $M=1$, $\vectg{\Sigma}(t, \vectg{\eta}) \coloneqq -k_1 \vectg{\Phi}(t, \vect{P}^{-1}\vectg{\eta})$;
    \item for $M\geq 2$, each $W \times W$ matrix block is defined by ratios of the system gains defined as $c_l \coloneqq k_{l+1}/k_{l}$, according to 
    \begin{equation*}
    \resizebox{.42\textwidth}{!}{$
    [\vectg{\Sigma}_{ij}]
    \coloneqq
     \begin{cases}
        -c_{M-1} \vect{I}_W & \text{, if } \: i=1\\
        (c_{M-i+1} - c_{M-i}) \vect{I}_W & \text{, if } \: 2 \leq i \leq j < M \\
        c_{M-j} \vect{I}_W & \text{, if } \: i=j+1\\        
        c_{1}\vect{I}_W - k_1 \vectg{\Phi}(t,\vect{P}^{-1}\vectg{\eta}) & \text{, if } \: j=i=M \\
        \vect{0} & \text{, if } \: i>j+1
    \end{cases}$}.
\end{equation*}
\end{enumerate}

Notice that in the new coordinate system the time-varying matrix $\vectg{\Phi}(t,\vect{P}^{-1}\vectg{\eta})$ only appears in the last block of the matrix $\vectg{\Sigma}(t,\vectg{\eta})$. 

\vspace{-0.1cm}
Consider the Lyapunov function
\begin{equation}
    V \coloneqq \frac{1}{2}\|\vectg{\eta}\|^2 = \frac{1}{2} \sum_{m=1}^{M} \|\vectg{\eta}^{(M-m)}\|^2 .
    \label{eqn:lyapunov_candidate_nonlinear}
\end{equation}
Taking its time-derivative yields $\dot{V} = -\frac{1}{2}\vectg{\eta}^{\top}\vect{Q}(t, \vectg{\eta})\vectg{\eta}$, where the matrix $\vect{Q}(t, \vectg{\eta})$ is symmetric and defined according to
\begin{equation}
	\vect{Q}(t,\vectg{\eta}) \coloneqq -\left(\vectg{\Sigma}(t,\vectg{\eta}) + \vectg{\Sigma}^{\top}(t, \vectg{\eta})\right) \in \vect{S}^{WM}.
	\label{eqn:auxiliar_Q_equation_definition}
\end{equation}

For $M=1$, the matrix $\vect{Q}(t, \vectg{\eta})$ simplifies to $\vect{Q}(t, \vectg{\eta}) = 2 k_1 \vectg{\Phi}(t, \vect{P}^{-1}\vectg{\eta})$. Assuming there exists $\delta > 0$ such that for all $t \geq t_0$,  
$\vectg{\Phi}(t, \vect{P}^{-1}\vectg{\eta}) \succ \delta \vect{I}$, then $\dot{V} \leq -k_1 \delta \|\vectg{\eta}\|^2$. It also follows directly from the comparison lemma (\cite{Khalil}) that the origin is UGES with
\begin{equation}
    \|\vectg{\eta}(t)\| \leq \|\vectg{\eta}(t_0)\| e^{-\delta k_1 (t-t_0)}.
    \label{eqn:rate_of_convergence_M1}
\end{equation}
For $M \geq 2$, since $\vect{Q}(t, \vectg{\eta})$ is a symmetric matrix, it can also be decomposed according to
\begin{equation*}
    \vect{Q}(t, \vectg{\eta}) 
    \coloneqq 
    \begin{bmatrix}
    	\vectg{\Upsilon}      & \vectg{\Gamma} \\
        \vectg{\Gamma}^{\top} & \vectg{\Delta}(t, \vectg{\eta}) 
    \end{bmatrix},
\end{equation*}
where $\vectg{\Delta}(t, \vectg{\eta}) = 2 \left(k_1\vectg{\Phi}(t, \vect{P}^{-1}\vectg{\eta}) - k_2/k_1 \: \vect{I}_W\right) \in \vect{S}^{W}$, and the matrices $\vect{\Upsilon} \in \mathbb{S}^{(M-1)W}$ and $\vect{\Gamma} \in \mathbb{R}^{(M-1)W \times W}$ can be obtained from \eqref{eqn:auxiliar_Q_equation_definition}. Assume there is $\beta>0$ for all $t\geq t_0$, and the system gains are appropriately chosen such that:
\begin{enumerate}[label=(\roman*)]
    \item the block matrix $\vectg{\Delta}(t, \vectg{\eta})$ satisfies $\vectg{\Delta}(t, \vectg{\eta}) \succ \beta\vect{I}_W$;
    \item the LMI given by
        \begin{equation}
    	\bar{\vect{Q}}
    	\coloneqq
    	\begin{bmatrix}
    		\vectg{\Upsilon} & \vectg{\Gamma} \\
    		\vectg{\Gamma}^{\top} & \beta\vect{I}_W
    	\end{bmatrix} \succ 0
    	\label{eqn:appendix:LMI_with_constant_term}
    \end{equation}
    is satisfied.
\end{enumerate}
Condition (i) is satisfied if inequality \eqref{eqn:condition_on_the_phi_matrix} is satisfied with $\delta = \beta/2$. In practice, and for design purposes, the LMI given by \eqref{eqn:appendix:LMI_with_constant_term} can also be re-written according to \eqref{eqn:LMI_for_theorem_1} and \eqref{eqn:gains_ratio_definition_theorem}, where the identity matrices are omitted, as they do not change the definiteness properties of $\bar{\vect{Q}}$. When conditions (i) and (ii) hold, it follows that $\vect{Q}(t, \vectg{\eta}) - \bar{\vect{Q}} \succeq 0$, and the time-derivative of the Lyapunov function verifies $\dot{V} = -\frac{1}{2}\vectg{\eta}^{\top} \vect{Q}(t, \vectg{\eta})\vectg{\eta} \leq -\frac{1}{2}\vectg{\eta}^{\top} \Bar{\vect{Q}}\vectg{\eta} \leq -\frac{1}{2}\lambda_{\min}(\Bar{\vect{Q}}) \| \vectg{\eta} \|^2$. From application of the comparison lemma (\cite{Khalil}), we can conclude that the system origin is UGES with
\begin{equation}
     \|\vectg{\eta}(t)\| \leq  \|\vectg{\eta}(t_0)\| e^{-\frac{1}{2} \lambda_{\min}(\Bar{\vect{Q}})(t-t_0)}, 
     \label{eqn:rate_of_convergence_M2}
\end{equation}
thus concluding the proof. \qed

\section{Proof of \autoref{lemma:ISS_stability_of_nonlinear_system}}
\label{appendix:proof_lemma_ISS_stability_of_nonlinear_system}
\textbf{Proof.} Consider the system \eqref{eqn:closed_loop_nonlinear_observer_canonical_form_with_input} and the similarity transformation \eqref{eqn:similarity_transformation} proposed in  \autoref{appendix:proof_theorem_stability_of_nonlinear_system}. The new system dynamics are given by $\dot{\vectg{\eta}} = \vectg{\Sigma}(t, \vectg{\eta})\vectg{\eta} + \vect{P}(\vect{B} \otimes \vect{u})$, which can be further simplified into $\dot{\vectg{\eta}} = \vectg{\Sigma}(t, \vectg{\eta})\vectg{\eta} + \frac{1}{k_M} \vect{B} \otimes \vect{u}$, with $\vect{B} = [1 \, \vect{0}^{\top}]^{\top} \in \mathbb{R}^{M}$. Consider the Lyapunov function \eqref{eqn:lyapunov_candidate_nonlinear}. Following the arguments presented in \autoref{appendix:proof_theorem_stability_of_nonlinear_system}, it follows directly that
\begin{equation*}
\begin{split}
    \dot{V} &= -\frac{1}{2}\vectg{\eta}^{\top} \vect{Q}(t, \vectg{\eta})\vectg{\eta} + \frac{1}{k_M} \vectg{\eta}^{\top}(\vect{B} \otimes \vect{u})\\
    &\leq -\frac{1}{2}\vectg{\eta}^{\top} \Bar{\vect{Q}}\vectg{\eta}  + \frac{1}{k_M} \vectg{\eta}^{\top}(\vect{B} \otimes \vect{u})\\
    &\leq -\frac{(1-\theta)}{2}\vectg{\eta}^{\top} \Bar{\vect{Q}}\vectg{\eta} \text{ for all } \|\vectg{\eta}\| \geq \frac{2}{\theta k_M \lambda_{\min}(\bar{\vect{Q}})}\|\vect{u}\|,
\end{split}
\end{equation*}
where $0 < \theta < 1$. Hence the system is ISS with respect to input $\vect{u}$ (\cite{Khalil}), provided that assertions (i) and (ii) from \autoref{theorem:stability_of_nonlinear_system} are satisfied. \qed

\section{Proof of \autoref{lemma:lemma_distributed_consensus_observer_ISS}}
\label{appendix:proof_lemma_distributed_consensus_observer_ISS}
\textbf{Proof.} Consider system \eqref{eqn:general_system_dynamics_for_observer} with a bounded input such that $\dot{\vect{x}}^{(M-1)} = \vect{u} \in \mathbb{R}^{K}$. The closed-loop observer error dynamics given by \eqref{eqn:total_closed_loop_observer_error_dynamics} will be affected by the new system input, according to $\dot{\tilde{\vect{x}}} = \vectg{\Xi}(t, \tilde{\vect{x}}) \tilde{\vect{x}} + \vect{B} \otimes \vect{u}^{\ast}$, where $\vect{u}^{\ast} \coloneqq \vect{1}_{N} \otimes \vect{u} \in \mathbb{R}^{KN}$ and $\vect{B} = [\vect{0}^{\top} \, 1]^{\top} \in \mathbb{R}^{M}$. By direct application of \autoref{lemma:ISS_stability_of_nonlinear_system}, and following the arguments presented in the proof of \autoref{theorem:theorem_distributed_consensus_observer}, it is trivial to conclude that the system is ISS with respect to input $\vect{u}$, provided that conditions (i), (ii), and (iii) from \autoref{theorem:theorem_distributed_consensus_observer} are satisfied. \qed